\documentclass{elsart}%[epj,numbook,nopacs]{svjour}

\usepackage{a4}
\usepackage{graphicx}
\usepackage{times}
\usepackage{amssymb,amsfonts,amsmath}   %AMS packages
\journal{Nuclear Physics B}
\begin{document}
 \begin{frontmatter}

\vspace{-2.5cm}
\hfill DESY 08-151 \\
\hfill Edinburgh 2008/39 \\[0.4cm]

  \title{
  Non-Perturbative Renormalization of Three-Quark Operators}
  \author[R]{Meinulf G\"ockeler},
  \author[E]{Roger Horsley},
  \author[R]{Thomas Kaltenbrunner},
  \author[DESY]{Yoshifumi Nakamura},
  \author[DESY]{Dirk Pleiter},
  \author[L]{Paul E. L. Rakow},
  \author[R]{Andreas Sch\"afer},
  \author[R,DESY]{Gerrit Schierholz},
  \author[Zuse]{Hinnerk St\"uben},
  \author[R]{Nikolaus Warkentin}, and
  \author[E]{James M. Zanotti} 
  \begin{center} (QCDSF/UKQCD Collaborations) \end{center}

  \address[R]{Institut f\"ur Theoretische Physik, Universit\"at Regensburg, 93040 Regensburg, Germany}
  \address[E]{School of Physics and Astronomy, University of Edinburgh, Edinburgh EH9 3JZ, UK}
  \address[DESY]{Deutsches Elektronen-Synchrotron DESY and John von Neumann Institut f\"ur Computing NIC, 15738 Zeuthen, Germany}
  \address[L]{Theoretical Physics Division, Department of Mathematical Sciences, University of Liverpool, Liverpool L69 3BX, UK}
  \address[Zuse]{Konrad-Zuse-Zentrum f\"ur Informationstechnik Berlin, 14195 Berlin, Germany}

\date{}

  \begin{abstract}
   High luminosity accelerators have greatly increased the interest in semi-exclusive and exclusive reactions involving nucleons. The relevant theoretical information is contained in the nucleon wavefunction and can be parametrized by moments of the nucleon distribution amplitudes, which in turn are linked to matrix elements of local three-quark operators. These can be calculated from first principles in lattice QCD. Defining an RI-MOM renormalization scheme, we renormalize three-quark operators corresponding to low moments non-perturbatively and take special care of the operator mixing. After performing a scheme matching and a conversion of the renormalization scale we quote our final results in the $\overline{\text{MS}}$ scheme at $\mu=2\,\text{GeV}$.
  \end{abstract}

 \end{frontmatter}

\section{Introduction}
\label{intro}
Distribution amplitudes play an essential role in the investigation of the internal nuclear structure. Exclusive high-energy processes can be factorized into hard and soft subprocesses, where the hard subprocess can be evaluated perturbatively and is characteristic for the reaction in question. The soft subprocess, described by the nucleon distribution amplitude $\Phi(x_i)$, contains the information about the distribution of the three valence quark momentum fractions $x_i$ inside the nucleon \cite{Lepage:1979za,Lepage:1980fj,Chernyak:1984bm,King:1986wi,Braun:2006hz}.

The great interest in this quantity stems from its importance for, e.g., the calculation of the electromagnetic form factors of the nucleon and their scaling behavior. These form factors describe a nucleon absorbing a virtual photon of squared momentum $-Q^2$ while remaining intact. According to \cite{Lepage:1980fj}, 
for $Q^2 \to \infty$, the magnetic form factor of the nucleon can be written as a convolution of three amplitudes: first, the distribution amplitude $\Phi$ for finding the nucleon in the valence state with the three quarks having definite momentum fractions $x_i$, second, the hard scattering kernel $T_H$, which describes one of the three quarks absorbing the photon, and finally the complex conjugate of $\Phi$ that gives the amplitude for the outgoing quarks to form a nucleon again:
\begin{align}
G_M(Q^2)=\int_0^1[dx] \int_0^1[dy] &\Phi^*(y_i,\tilde Q_y) T_H(x_i,y_i,Q) \Phi(x_i,\tilde Q_x) \, \left( 1+ \mathcal{O}(m^2_N/Q^2) \right).
\label{eq_GM}
\end{align}
Here $[dx]\equiv dx_1dx_2dx_3 \delta(1-\sum_i x_i)$, $\tilde Q_x\equiv \text{min}_i(x_iQ)$ and $m_N$ denotes the nucleon mass.

Distribution amplitudes are genuinely \hyphenation{non-per-tur-ba-tive} non-perturbative quantities. Hence they are inaccessible for perturbation theory and must be calculated by other means, e.g., by lattice QCD. After a pioneering study in the late 1980s \cite{Martinelli:1988xs} only recently the first quantitative results from lattice QCD have been published \cite{Gockeler:2007qs,Gockeler:2008xv,Niko_Forthcoming}.
After performing an expansion near the lightcone, moments of the nucleon distribution amplitudes are expressed in terms of matrix elements of local three-quark operators that are evaluated between a nucleon state and the vacuum. These three-quark operators typically consist of a combination of covariant derivatives acting on the three quark fields, which are located at a common space-time coordinate $x$. Furthermore, the operators for the nucleon are color singlets and have isospin $1/2$.

As the three-quark operators pick up radiative corrections and are subject to mixing with other operators, their renormalization is a vital ingredient for any lattice calculation. In a previous paper \cite{Kaltenbrunner:2008pb} we have derived irreducible multiplets of three-quark operators with respect to the spinorial hypercubic group $\overline{\text{H}(4)}$ and have discussed the mixing properties of these operators in detail. The present work will focus on the non-perturbative renormalization of these three-quark operators in the isospin-$1/2$ sector.

 The paper is organized as follows. In the first section, we will introduce an RI-MOM renormalization scheme which will be implemented in our lattice calculations. Then we will explain how to perform a perturbative scheme matching to $\overline{\text{MS}}$ and derive the anomalous dimensions of the operators in question. The following two sections will focus on a discussion of our results for the renormalization matrices, which are finally quoted at $\mu=2\,\text{GeV}$ in the $\overline{\text{MS}}$ scheme. In the final section we will demonstrate how to renormalize moments of the nucleon distribution amplitude and present the results of consistency checks between the renormalized moments.

\section{Lattice Renormalization}
\label{sec:1}
 Our main aim is the derivation of non-perturbative renormalization coefficients for three-quark operators in lattice QCD and their application to the renormalization of moments of the nucleon distribution amplitude. As discussed in the introduction, a subsequent convolution with the hard scattering kernel leads then, e.g., to estimates for the electromagnetic form factors of the nucleon. Therefore the renormalization of the hard scattering kernel and the three-quark operators inside the nucleon distribution amplitudes must be carried out in a consistent way, i.e., the same renormalization scheme has to be applied. As the perturbative results for the hard subprocess are usually given in the $\overline{\text{MS}}$ scheme, the distribution amplitudes must also be renormalized in this scheme.

 On the lattice, however, it is not possible to implement the $\overline{\text{MS}}$ scheme. Therefore one introduces the RI-MOM renormalization scheme \cite{Martinelli:1994ty}. It is applicable both on the lattice and in the continuum. Then one first renormalizes the matrix elements on the lattice within this scheme. Afterwards one applies continuum perturbation theory to calculate a matching function between both schemes and extracts non-perturbatively renormalized lattice operators in the $\overline{\text{MS}}$ scheme. In this section we will discuss the setup and implementation of the lattice renormalization.

\subsection{The Three-Quark Operators}
 We are interested in three-quark operators for the nucleon. Choosing the flavors $u$, $u$ and $d$ for definiteness, their general form is
 \begin{align}
  \mathcal{O}^{uud}_i(x) =\,& \epsilon_{c_1' c_2' c_3'}\,
  T^{(i)}_{\mu_1 \dots \mu_m \nu_1 \dots \nu_n \lambda_1 \dots \lambda_l \alpha' \beta' \gamma'} \, \left( D_{\mu_1} \dots D_{\mu_m} u(x)_{\alpha'}\right)_{c_1'} \nonumber\\
  &\cdot \left(D_{\nu_1} \dots D_{\nu_n} u(x)_{\beta'}\right)_{c_2'} 
  \, \left(D_{\lambda_1} \dots D_{\lambda_l} d(x)_{\gamma'}\right)_{c_3'},
  \label{el3qOp}
 \end{align}
 with analogous definitions for $\mathcal{O}^{udu}_i(x)$ and $\mathcal{O}^{duu}_i(x)$. 
After projection onto isospin 1/2 one ends up with operators for 
the proton, and subsequent exchange of the $u$ and $d$ fields leads to 
neutron operators.
The color indices $c_j'$ will be suppressed when not required. The coefficient tensor $T^{(i)}$ determines the symmetry structure of the spinor indices $\alpha'$, $\beta'$, $\gamma'$ and of the space-time indices $\mu_j$, $\nu_j$, $\lambda_j$ of the covariant derivatives $D$ contributing to the operator $\mathcal{O}_i$.

 In \cite{Kaltenbrunner:2008pb} we have derived multiplets of three-quark operators that transform irreducibly under the spinorial symmetry group of the hypercubic lattice $\overline{\text{H}(4)}$. 
These operators reduce the problem of mixing under renormalization to 
mixing among multiplets belonging to equivalent representations of 
$\overline{\text{H}(4)}$.
Hence they define our choice for the coefficient tensors $T^{(i)}$. Appendix B of \cite{Kaltenbrunner:2008pb} contains all linearly independent sets of potentially mixing isospin-1/2 three-quark operators with leading twist and up to two derivatives. They provide the basis for the (non-perturbative) renormalization. Operators with total derivatives $\partial_\mu$ are automatically taken into account due to the identity $\partial_\mu(fgh) =(D_\mu f)gh +f (D_\mu g)h + fg(D_\mu h)$ for any color-singlet operator made of the three quark fields $f$, $g$ and $h$. Note however that this continuum relation holds only up to discretization errors on the lattice.

 Let us finally give an example of a typical three-quark operator belonging to an irreducibly transforming multiplet. We have, e.g., the following operator with two derivatives:
 \begin{align}
  \mathcal{O}_{fg17}^{(4),\text{MA}} = \frac{5i}{8\sqrt{3}} &\left( \frac{3}{5} {(D\sigma)^{\{0}}_{\{ \dot 0} u^{0} {(D\sigma)^0}_{\dot 0 \}} d^{0}  u^{0 \}} \right. - \frac{3}{5} {(D\sigma)^{\{0}}_{\{ \dot 0} d^{0} {(D\sigma)^0}_{\dot 0 \}} u^{0}  u^{0 \}}  \nonumber\\
                          &- {(D\sigma)^{\{1}}_{\{ \dot 0} u^{1} {(D\sigma)^1}_{\dot 0 \}} d^{1}  u^{0 \}} + {(D\sigma)^{\{1}}_{\{ \dot 0} d^{1} {(D\sigma)^1}_{\dot 0 \}} u^{1}  u^{0 \}} \nonumber\\
                          & - 2 \cdot {(D\sigma)^{\{0} }_{\{ \dot 1} u^{1} {(D\sigma)^0}_{\dot 1 \}} d^{1}   u^{0 \}}  \left. + 2 \cdot {(D\sigma)^{\{0} }_{\{ \dot 1} d^{1} {(D\sigma)^0}_{\dot 1 \}} u^{1}   u^{0 \}} \right).
 \end{align}
 The superscript MA (standing for ``mixed antisymmetric'') indicates that the operator has isospin 1/2 while the subscript $fg$ means that the derivatives act on the first and second quark, compare \cite{Kaltenbrunner:2008pb}. Pauli matrices $\sigma_\mu$ are used to contract the covariant derivatives. The curly braces denote independent total symmetrization in the dotted and undotted indices of the Weyl representation.

\subsection{Calculational Method}
 We introduce a correlation function for the non-perturbative renormalization of the three-quark operators by contracting the operators with three external quark sources, namely $\overline{u}(z_1)$, $\overline{u}(z_2)$ and $\overline{d}(z_3)$:
 \begin{align}
  \langle \overline{u}(z_1)_{\alpha c_1} \overline{u}(z_2)_{\beta c_2} \overline{d}(z_3)_{\gamma c_3} \mathcal{O}_i(x) \rangle.
  \label{eq_local3qOMatrixElement}
 \end{align}
 We have two $u$-quark lines and one $d$-line running from the three space-time coordinates $z_i$ into a vertex at $x$. There they are connected according to the coefficient tensor $T^{(i)}$.

 We proceed in analogy to the case of quark-antiquark operators \cite{Martinelli:1994ty,Gockeler:1998ye}. We impose fixed momentum on the external quark lines and evaluate the correlation function in momentum space. The result is a four-point function $G(p_1, p_2, p_3)^{(i)}_{\alpha \beta \gamma}$ that depends on the momentum of the three external quark lines and carries three spinor indices:
 \begin{align}
  G(p_1,p_2,p_3)^{(i)}_{\alpha \beta \gamma} =  \frac{1}{V} \sum_{x,z_1,z_2,z_3} \exp(+ i p_1 \cdot z_1 + i p_2 \cdot z_2 + i p_3 \cdot z_3) \qquad\qquad \nonumber\\
      \cdot \exp(-i(p_1+p_2+p_3)\cdot x) \,\epsilon_{c_1 c_2 c_3}\,
   \langle \overline{u}(z_1)_{\alpha c_1} \overline{u}(z_2)_{\beta c_2} \overline{d}(z_3)_{\gamma c_3} \mathcal{O}_i(x) \rangle.
 \label{eq_G}
 \end{align}
 This four-point function is obtained as the ensemble average of quark field contractions on the individual gauge configurations. Let us denote such a contraction on a single configuration by the brackets $[\dots]$. For $\mathcal{O}_i = \mathcal{O}^{uud}_i$ we then have, e.g.:
 \begin{align}
  G&(p_1,p_2,p_3)^{(i)}_{\alpha \beta \gamma} = 
  \frac{1}{N}\sum_{\text{config.}} \epsilon_{c_1' c_2' c_3'} \, \epsilon_{c_1 c_2 c_3} \,T^{(i)}_{\mu_1 \dots \mu_m \nu_1 \dots \nu_n \lambda_1 \dots \lambda_l \alpha' \beta' \gamma'} \nonumber\\
  &\cdot \frac{1}{V} \sum_{x,z_1,z_2,z_3} \exp(-i(p_1+p_2+p_3)\cdot x) \, \exp(+ i p_1 \cdot z_1 + i p_2 \cdot z_2 + i p_3 \cdot z_3) \nonumber\\
  &\cdot \left( \left(D^x_{\mu_1} \dots D^x_{\mu_m}\right)_{c_1' c_1''} [u(x)_{\alpha',c_1''} \overline{u}(z_1)_{\alpha,c_1}] \cdot \left( D^x_{\nu_1} \dots D^x_{\nu_n}\right)_{c_2' c_2''} [u(x)_{\beta',c_2''} \overline{u}(z_2)_{\beta,c_2}] \right. \nonumber\\
  &+ \left. \left(D^x_{\mu_1} \dots D^x_{\mu_m}\right)_{c_2' c_2''} [u(x)_{\beta',c_2''} \overline{u}(z_1)_{\alpha,c_1}] \cdot \left(D^x_{\nu_1} \dots D^x_{\nu_n}\right)_{c_1' c_1''} [u(x)_{\alpha', c_1''} \overline{u}(z_2)_{\beta,c_2}] \right) \nonumber\\
 &\cdot \left(D^x_{\lambda_1} \dots D^x_{\lambda_l} \right)_{c_3' c_3''} [d(x)_{\gamma',c_3''} \overline{d}(z_3)_{\gamma,c_3}],
 \end{align}
 where $N$ is the number of gauge field configurations. Analogous expressions hold for the four-point functions of the operators $\mathcal{O}^{udu}_i$ and $\mathcal{O}^{duu}_i$ as well as for the operators with definite isospin.

 Let us now define
 \begin{align}
  K(x,p)_{\alpha_1 c_1, \alpha_2 c_2} \equiv \sum_{z} \exp(i p\cdot z) [u(x)_{\alpha_1, c_1} \overline{u}(z)_{\alpha_2,c_2}].
  \label{eq_K}
 \end{align}
 This quantity can be determined on every single gauge configuration by inverting the massive Dirac operator $M$ of the action on a 
momentum source~\cite{Gockeler:1998ye}:
 \begin{align}
  \sum_{x,\alpha_1,c_1} M_{\alpha_0 c_0, \alpha_1 c_1}(y,x)  K(x,p)_{\alpha_1 c_1, \alpha_2 c_2} = \exp(i p\cdot y) \delta_{\alpha_0 \alpha_2} \delta_{c_0 c_2}.
 \end{align}
 Rewriting the above four-point function $G$ in terms of $K$ provides insight into the lattice implementation:
 \begin{align}
  G&(p_1,p_2,p_3)^{(i)}_{\alpha \beta \gamma} = \frac{1}{N}\sum_{\text{config.}} \epsilon_{c_1' c_2' c_3'} \, \epsilon_{c_1 c_2 c_3} \, T^{(i)}_{\mu_1 \dots \mu_m \nu_1 \dots \nu_n \lambda_1 \dots \lambda_l \alpha' \beta' \gamma'} \nonumber\\
  &\cdot \frac{1}{V} \sum_{x} \exp(-i(p_1+p_2+p_3)\cdot x)\, \left(D^x_{\lambda_1} \dots D^x_{\lambda_l} \right)_{c_3' c_3''} K(x,p_3)^d_{\gamma' c_3'', \gamma c_3} \nonumber\\
  &\cdot \left( \left(D^x_{\mu_1} \dots D^x_{\mu_m}\right)_{c_1' c_1''} K(x,p_1)^u_{\alpha' c_1'', \alpha c_1} \cdot \left( D^x_{\nu_1} \dots D^x_{\nu_n} \right)_{c_2' c_2''} K(x,p_2)^u_{\beta' c_2'', \beta c_2} \right. \nonumber\\
  &+ \left. \left(D^x_{\mu_1} \dots D^x_{\mu_m} \right)_{c_2' c_2''} K(x,p_1)^u_{\beta' c_2'', \alpha c_1} \cdot \left(D^x_{\nu_1} \dots D^x_{\nu_n} \right)_{c_1' c_1''} K(x,p_2)^u_{\alpha' c_1'', \beta c_2} \right).
 \end{align}
 The most expensive step in the calculation is the evaluation and symmetrization of the spin-color combinations which is, naively speaking, due to the existence of an additional quark and antiquark field with 12 spin-color indices, by a factor of $12\times12=144$ more expensive than for mesonic operators. So, special care has to be taken when implementing these contractions. As the correlation function is not gauge invariant, all configurations are gauge fixed to Landau gauge.

 In a final step we amputate the external quark lines of the four-point function $G^{(i)}$ to arrive at the three-quark vertex $\Gamma^{(i)}$:
 \begin{align}
  G(p_1,p_2,p_3)^{(i)}_{\alpha \beta \gamma} = \Gamma(p_1,p_2,p_3)^{(i)}_{\alpha' \beta' \gamma'} S(p_1)_{\alpha' \alpha} S(p_2)_{\beta' \beta} S(p_3)_{\gamma' \gamma},
  \label{eq_Gamma}
 \end{align}
 where the quark propagators are defined by
 \begin{align}
  S(p)_{\alpha_1 \alpha_2} = \frac{1}{V} \sum_{x,y} \langle u(x)_{\alpha_1} \overline{u}(y)_{\alpha_2} \rangle \exp(-ip\cdot (x-y)).
 \end{align}
 Note that we can reuse the quantity $K(x,p)$ introduced in (\ref{eq_K}) to calculate the propagator $S(p)$:
 \begin{align}
  S(p)_{\alpha_1 \alpha_2} = \frac{1}{V} \frac{1}{N}\sum_{\text{config.}} \sum_x K(x,p)_{\alpha_1 \alpha_2} \exp(-ip\cdot x).
 \end{align}
 In the following section we will define our renormalization scheme based on the vertex $\Gamma^{(i)}$.

\subsection{An RI-MOM Renormalization Scheme}
 \label{subsec:1}
 We want to introduce a renormalization scheme that is applicable on the lattice and in the continuum. For quark-antiquark operators such a scheme has been proposed in \cite{Martinelli:1994ty} and is widely known as the RI-MOM scheme. In \cite{Aoki:2006ib} it was also used for the renormalization of proton decay matrix elements. To study the mixing of our three-quark operators with up to two derivatives, we have slightly modified this approach. In the following, we will set up our modified RI-MOM renormalization scheme. As, in general, mixing is a central issue of the renormalization, we will consider mixing matrices explicitly from the very beginning.

 The renormalized counterpart of a general regularized operator $\mathcal{O}_i$ is given by
 \begin{align}
  \mathcal{O}^{\text{ren}}_i = Z_{ij} \mathcal{O}_j.
  \label{eq_ZO}
 \end{align}
 Here $Z_{ij}$ is the renormalization matrix and operator mixing shows up in non-vanishing off-diagonal elements of $Z$.

 We will define the renormalization matrix $Z^{\text{mRI}}$ for three-quark operators by projections of the lattice-regularized three-quark vertex $\Gamma$ introduced in eq. (\ref{eq_Gamma}). In the following we will distinguish between the tree-level, the lattice regularized (lattice spacing $a$) and the renormalized vertices $\Gamma^{\text{tree}}_i$, $\Gamma^{\text{latt}}_i$ and $\Gamma^{\text{mRI}}_i$, respectively. Let us now introduce a set of projectors $P_k$ in spinor space that fulfill the following orthogonality condition with the tree-level vertices:
 \begin{align}
  P_k \Gamma^{\text{tree}}_i(p_1,p_2,p_3) = \delta_{ki}.
  \label{eq_Projectors}
 \end{align}
 At some renormalization scale $\mu$ fixed by the mean squares of the three external quark momenta we then require the renormalized three-quark vertex to fulfill the same equation. This yields the renormalization condition
 \begin{align}
  P_k \Gamma^{\text{mRI}}_i(p_1,p_2,p_3;\mu)\vert_{\mu^2=\sum_i p_i^2/3} = \delta_{ki}.
 \end{align}
 With $\Gamma^{\text{mRI}}_i = Z_{ij}^{\Gamma,\text{mRI}} \Gamma^{\text{latt}}_j$ we can introduce the auxiliary variable $Z_{ij}^{\Gamma,\text{mRI}}$ for the renormalization of the vertex:
 \begin{align}
  (Z_{\Gamma,\text{mRI}}^{-1})_{ij}(\mu) &= P_j  \Gamma^{\text{latt}}_i(p_1,p_2,p_3)\vert_{\mu^2=\sum_i p_i^2/3}
  \label{eq_ZGamma}
 \end{align}

 In any scheme $Z^{\Gamma}$ is related to the renormalization matrix of the three-quark operators $\mathcal{O}$ by the quark field renormalization $Z_q$. To compensate for the amputated quark legs one needs a factor of $Z_q^{1/2}$ for each of them:
 \begin{align}
  Z_{ij}(\mu) =  Z_q(p_1)^{1/2} \, Z_q(p_2)^{1/2} \, Z_q(p_3)^{1/2} \cdot Z^{\Gamma}_{ij}(\mu).
  \label{eq_Z}
 \end{align}

 This fixes the three-quark operator renormalization matrix in the RI-MOM scheme:
 \begin{align}
  (Z_{\text{mRI}}^{-1})_{ij}(\mu) &=& Z^{\text{RI}^\prime}_q(p_1)^{-1/2}\, 
    Z^{\text{RI}^\prime}_q(p_2)^{-1/2} \, Z^{\text{RI}^\prime}_q(p_3)^{-1/2} \cdot P_j  
    \Gamma^{\text{latt}}_i(p_k)\vert_{\mu^2=\sum_k p_k^2/3}. &
  \label{eq_RenMaster}
 \end{align}
 As usual we determine the factors $Z_q$ in the $\text{RI}^\prime$
scheme \cite{Martinelli:1994ty}. The peculiarity of the mRI scheme lies in the special definition of the projectors for the three-quark vertex $\Gamma$ and will be explained in the next subsection.

 \subsection{Choice of the Projectors}
 \label{Sec:3}
 In order to determine the RI-MOM renormalization matrix $Z$ we have introduced but not yet defined a set of projectors $P_j$. The only restriction is given by eq. (\ref{eq_Projectors}). 

 We now turn to vector notation. $\Gamma_i$ is a tensor of rank three in spinor-space and can be interpreted as a vector $v_{\Gamma_i}$ of dimension $4^3$. Then we can interpret the projectors $P_k$ as orthogonal projections onto vectors $v_{P_k}$ of the same dimension:
 \begin{align}
  P_k \Gamma_i \equiv \langle v_{P_k}, v_{\Gamma_i} \rangle,
 \end{align}
 where
 \begin{align}
  \langle v_1, v_2 \rangle \equiv \sum_j (v_1)^*_j \cdot (v_2)_j.
 \end{align}

 The task is now to construct a set of vectors $v_{P_k}$ that fulfills the normalization condition eq. (\ref{eq_Projectors}), which reads in vector notation
 \begin{align}
  \langle v_{P_k}, v_{\Gamma^{\text{tree}}_i} \rangle = \delta_{ki}.
 \end{align}
 We choose the vectors $v_{P_k}$ as follows. We start with an auxiliary vector
 \begin{align}
  v'_{P_k} = v_{\Gamma^{\text{tree}}_k},
 \end{align}
 and project it onto the orthogonal complement of the space spanned by the vectors $v_{\Gamma^{\text{tree}}_j}$, $j \ne k$. This results in an altered vector $v''_{P_k}$. Taking care of the normalization we finally define
 \begin{align}
  v_{P_k} = \frac{1}{\langle v_{\Gamma^{\text{tree}}_k}, v''_{P_k} \rangle} v''_{P_k}.
  \label{eq_ProjectorVector}
 \end{align}
 By now all constituents of the mRI renormalization scheme are defined. We summarize the method by rewriting eq. (\ref{eq_RenMaster}) in vector notation:
 \begin{align}
  (Z_\text{mRI}^{-1})_{ij}(\mu) = Z_q^{\text{RI}^\prime}(p_1)^{-1/2}\, Z_q^{\text{RI}^\prime}(p_2)^{-1/2} \, Z_q^{\text{RI}^\prime}(p_3)^{-1/2} \cdot \langle v_{P_j} , v_{\Gamma^{\text{latt}}_i(p_k)} \rangle \vert_{\mu^2=\sum_k p_k^2/3}.
  \label{eq_RenScalarProduct}
 \end{align}
 Note that the above renormalization condition will in general depend on the geometry of the external momenta, i.e., the angles between the four-momenta $p_k$. In the end this dependence will be cancelled by the scheme matching. We want to stress furthermore that, due to its general structure, the method is not limited to the case of three-quark operators and four-point functions discussed in this paper, but is applicable to the general class of $n$-point functions.

\section{Perturbative Calculations}
 In the introduction we have emphasized the importance of renormalized three-quark operators for nucleon distribution amplitudes. To calculate observables, both the distribution amplitude and the hard scattering kernel must be given in the same renormalization scheme.
 In the following we will explain, how a matching of the RI-MOM scheme to $\overline{\text{MS}}$ can be achieved with the help of continuum perturbation theory. Moreover we will study the dependence of the renormalization coefficients on the renormalization scale $\mu$. As for the lattice computation, also here all calculations have to be carried out in the Landau gauge.

\subsection{Scheme Matching to $\overline{\text{MS}}$}
 Let us start by writing eq. (\ref{eq_ZO}) explicitly in both renormalization schemes:
 \begin{align}
  \mathcal{O}^{\text{mRI}}_i = Z^{\text{mRI}}_{ij} \mathcal{O}_j, \nonumber\\
  \mathcal{O}^{\overline{\text{MS}}}_i = Z^{\overline{\text{MS}}}_{ij} \mathcal{O}_j,
 \end{align}
 with $\mathcal{O}^{\text{mRI}}_i$ and $\mathcal{O}^{\overline{\text{MS}}}_i$ denoting the renormalized operators in the modified RI and the $\overline{\text{MS}}$ scheme, respectively. If we introduce the scheme matching matrix
 \begin{align}
  Z^{\overline{\text{MS}}\leftarrow \text{mRI}}_{ij} = Z^{\overline{\text{MS}}}_{ik} \left((Z^{\text{mRI}})^{-1}\right)_{kj},
  \label{eq_SchemeMatching}
 \end{align}
 we get a relation between the operators renormalized in the $\overline{\text{MS}}$ and mRI schemes:
 \begin{align}
  \mathcal{O}^{\overline{\text{MS}}}_i = Z^{\overline{\text{MS}}\leftarrow \text{mRI}}_{ij} \mathcal{O}^{\text{mRI}}_j.
 \end{align}
 In the following we derive the matching functions with dimensional regularization in continuum perturbation theory.

 To this end we proceed with a one-loop perturbative expansion of $Z^{\overline{\text{MS}}}$ and $Z^{\text{mRI}}$. Up to $\mathcal{O}(\epsilon)$ the renormalization matrix reads in the mRI scheme:
 \begin{align}
  Z^{\text{mRI}}_{ij} &= \delta_{ij} + \frac{\alpha_s(\mu)}{4\pi} \left( Z^{\text{mRI}}_0 \right)_{ij} + \frac{\alpha_s(\mu)}{4\pi} \frac{1}{\bar\epsilon} \left( Z^{\text{mRI}}_1 \right)_{ij}. \label{eq_SM_ZRI}
 \end{align}
 Here we have used the renormalized strong coupling $\alpha_s(\mu)=g_R(\mu)^2/4\pi$ and adopted the following conventions for the dimensional regularization:
 \begin{align}
  \frac{1}{\bar \epsilon} = &\,\frac{1}{\epsilon} + \frac{1}{2} \log 4\pi - \frac{1}{2} \gamma_E, \\
  \epsilon = &\,4-d.
 \end{align}
 In the numerical evaluation of $\alpha_s$ we use $\Lambda^{\overline{\text{MS}}}= 261\,\text{MeV}$ \cite{Gockeler:2005rv}.
 Comparing eq. (\ref{eq_SM_ZRI}) with the analogous expression for $Z^{\overline{\text{MS}}}$ and noting that the conversion between both must be finite, the scheme matching matrix in first order becomes
 \begin{align}
  Z^{\overline{\text{MS}}\leftarrow \text{mRI}}_{ij}(\mu) = \delta_{ij} - \frac{\alpha_s(\mu)}{4\pi} \left( Z^{\text{mRI}}_0 \right)_{ij} +\mathcal{O}(\alpha_s^2).
  \label{eq_ZMSRI}
 \end{align}
 This means that $Z^{\overline{\text{MS}}\leftarrow \text{mRI}}_{ij}$ can be derived from the perturbative expansion of $Z^{\text{mRI}}_{ij}$ alone. In the following section we will discuss this in more detail.

\subsection{Determination of $Z^{\overline{\text{MS}}\leftarrow \text{mRI}}$}
 According to eq. (\ref{eq_Z}) the renormalization of the three-quark operators consists of four parts:
 \begin{align}
  Z^{\text{mRI}}_{ij} = (Z_q^{{\text{RI}^\prime}})^{1/2}\, (Z_q^{{\text{RI}^\prime}})^{1/2}\, (Z_q^{{\text{RI}^\prime}})^{1/2} \cdot Z^{\Gamma,\text{mRI}}_{ij} .
 \end{align}
 The scheme matching can be performed independently for each renormalization factor so that
 \begin{align}
  Z^{\overline{\text{MS}}\leftarrow \text{mRI}}_{ij} = (Z_q^{\overline{\text{MS}}\leftarrow {\text{RI}^\prime}})^{1/2}\, (Z_q^{\overline{\text{MS}}\leftarrow {\text{RI}^\prime}})^{1/2}\, (Z_q^{\overline{\text{MS}}\leftarrow {\text{RI}^\prime}})^{1/2} \cdot Z^{\Gamma,\overline{\text{MS}}\leftarrow \text{mRI}}_{ij}.
  \label{eq:ZOpMatching}
 \end{align}
 In two-loop order the matching of the quark field renormalization reads \cite{Chetyrkin:1999pq}:
 \begin{align}
  Z_q^{\overline{\text{MS}}\leftarrow {\text{RI}^\prime}} = 1- \frac{\alpha_s}{4\pi} \frac{4 \xi}{3} + \Big( \frac{\alpha_s}{4\pi} \Big)^2 \Big(-\frac{49 \xi ^2}{18}+12 \zeta_3 \xi -26 \xi 
     +\frac{7}{3} n_f+12 \zeta_3-\frac{359}{9} \Big),
  \label{eq:ZqMatching}
 \end{align}
 where $\xi$ is the covariant gauge parameter, compare Appendix \ref{App:FeynmanDiagrams}.
 The scheme matching for the renormalization coefficient $Z^{\Gamma}$ of the three-quark vertex will be determined in one-loop order. In analogy to eq. (\ref{eq_ZMSRI}) we have
 \begin{align}
  Z^{\Gamma,\overline{\text{MS}}\leftarrow \text{mRI}}_{ij} &= \delta_{ij} - \frac{\alpha_s}{4\pi} \left( Z^{\Gamma, \text{mRI}}_0 \right)_{ij} + \mathcal{O}(\alpha_s^2).
  \label{eq_SM_ZOZGamma}
 \end{align}
 To evaluate this expression, the renormalization matrix for the three-quark vertex $\Gamma$ is needed in the mRI scheme. Therefore we perform a perturbative expansion of the dimensionally regularized three-quark vertices:
 \begin{align}
  \Gamma^{\text{dim}}_i = \Gamma_i^{\text{tree}} + \frac{\alpha_s(\mu)}{4\pi} \, \Gamma_{i,0}^{\text{dim}}(\mu,p_k) + \frac{\alpha_s(\mu)}{4\pi}\, \frac{1}{\bar\epsilon}\, \Gamma_{i,1}^{\text{dim}}(\mu,p_k).
  \label{eq_GammaDim}
 \end{align}
 If we apply the projectors introduced in eq. (\ref{eq_Projectors}) and make use of their linearity, we find
 \begin{align}
  Z^{\Gamma,\text{mRI}}_{ij}(\mu) = \phantom{\frac{1}{1}}\!\!\! \delta_{ij} - \frac{\alpha_s(\mu)}{4\pi} \langle v_{P_j} , v_{\Gamma_{i,0}^{\text{dim}}(\mu,p_k)} \rangle -\frac{\alpha_s(\mu)}{4\pi} \frac{1}{\bar\epsilon} \langle v_{P_j} , v_{\Gamma_{i,1}^{\text{dim}}(\mu,p_k)} \rangle.
 \end{align}
 Comparing with eq. (\ref{eq_SM_ZRI}) reveals the identities
 \begin{align}
  \left( Z^{\Gamma, \text{mRI}}_0 \right)_{ij} = - \langle v_{P_j} , v_{\Gamma^{\text{dim}}_{i,0}(\mu,p_k)} \rangle, \nonumber\\
  \left( Z^{\Gamma, \text{mRI}}_1 \right)_{ij} = - \langle v_{P_j} , v_{\Gamma^{\text{dim}}_{i,1}(\mu,p_k)} \rangle,
 \end{align}
 which have to be evaluated at $\mu^2=(p_1^2+p_2^2+p_3^2)/3$ for the momentum geometries used in the simulations. Inserting this into eq. (\ref{eq_SM_ZOZGamma}) yields the result for the scheme matching matrix of the vertex in first order:
 \begin{align}
  Z^{\Gamma,\overline{\text{MS}}\leftarrow \text{mRI}}_{ij}(\mu) = \delta_{ij} + \frac{\alpha_s(\mu)}{4\pi} \, \langle v_{P_j} , v_{\Gamma^{\text{dim}}_{i,0}(\mu,p_k)} \rangle + \mathcal{O}(\alpha_s^2).
  \label{eq_ZMSRI_Master}
 \end{align}
 Together with eqs. (\ref{eq:ZqMatching}) and (\ref{eq:ZOpMatching}) this determines our scheme matching for the three-quark operator.

 The perturbative calculation of $\Gamma^{\text{dim}}_{i,0}(\mu,p_k)$ for the different three-quark operators is carried out in Euclidean space-time with off-shell quarks and gluons in Landau gauge $\xi=0$. It results in lengthy expressions and not all occurring integrals over the Feynman parameters can be solved analytically in closed form. The final evaluation of the integrals over the Feynman parameters and the construction of the projectors $P_j$ as well as the evaluation of $\langle v_{P_j} , v_{\Gamma^{\text{dim}}_{i,0}(\mu,p_k)} \rangle$ were performed with Mathematica for all required momentum combinations.

 Note that it is important to exercise care when evaluating the terms of order $\mathcal{O}(\epsilon^0)$ in the three-quark vertex that define $\Gamma^{\text{dim}}_{i,0}$. Generally speaking one has to keep track of all possible terms in the Dirac structures proportional to $\epsilon$ that multiply a $\frac{1}{\bar \epsilon}$ divergence, since these produce additional contributions to the scheme matching matrix. We have used the following strategy to treat the continuation to $d$ dimensions. In a first step we have written our irreducible three-quark operators in four dimensions as linear combinations of the following basis operators:
 \begin{align}
  A^{\rho \lambda_i, \mu_i, \nu_i}_\tau =\, & \epsilon_{c_1 c_2 c_3} (D_{\lambda_1}\dots D_{\lambda_l} u_\alpha)_{c_1} (C \gamma_\rho \gamma_5)_{\alpha \beta} (D_{\mu_1} \dots D_{\mu_m} u_\beta)_{c_2} \nonumber\\
                                            &\times (D_{\nu_1} \dots D_{\nu_n} d_\tau)_{c_3}, \nonumber\\
  V^{\rho \lambda_i, \mu_i, \nu_i}_\tau =\, & \epsilon_{c_1 c_2 c_3} (D_{\lambda_1}\dots D_{\lambda_l} u_\alpha)_{c_1} (C \gamma_\rho)_{\alpha \beta} (D_{\mu_1} \dots D_{\mu_m} u_\beta)_{c_2} \nonumber\\
                                            &\times (D_{\nu_1} \dots D_{\nu_n} (\gamma_5 d)_\tau)_{c_3}, \nonumber\\
  W^{\rho \lambda_i, \mu_i, \nu_i}_\tau =\, & \epsilon_{c_1 c_2 c_3} (D_{\lambda_1}\dots D_{\lambda_l} u_\alpha)_{c_1} (C \gamma_\rho)_{\alpha \beta} (D_{\mu_1} \dots D_{\mu_m} d_\beta)_{c_2} \nonumber\\
    &\times (D_{\nu_1} \dots D_{\nu_n} (\gamma_5 u)_\tau)_{c_3}, \nonumber\\
  U^{\rho\mu \lambda_i, \mu_i, \nu_i}_\tau =\, & \epsilon_{c_1 c_2 c_3} (D_{\lambda_1}\dots D_{\lambda_l} u_\alpha)_{c_1} (C (-i) \sigma_{\rho \mu})_{\alpha \beta} (D_{\mu_1} \dots D_{\mu_m} u_\beta)_{c_2} \nonumber\\
     &\times (D_{\nu_1} \dots D_{\nu_n} (\gamma_5 d)_\tau)_{c_3}.
  \label{eq:ZusatzEpsBasis}
 \end{align}
 As usual $\alpha$, $\beta$, $\gamma$ and $\tau$ denote spinor indices, $\lambda_i$, $\mu_i$, $\nu_i$ as well as $\rho$ and $\mu$ are space-time indices. Note that the operator $W$ is equal to the operator $V$ up to the position of the down quark.
 For the special case of operators without derivatives we have also used
 \begin{align}
  \tilde U^{\rho\mu}_\tau =\, & \epsilon_{c_1 c_2 c_3} u_{\alpha c_1} (C (-i) \sigma_{\rho \mu})_{\alpha \beta} d_{\beta c_2} (\gamma_5 u)_{\tau c_3},
 \end{align}
 to access the operators of sub-leading twist.

 Then we have rewritten the three-quark vertices belonging to the above operator basis in terms of dimensionally regularized loop integrals. The corresponding Feynman diagrams consist of three quark lines and one gluon exchange, leading to three strings of gamma matrices in the associated amplitudes. In the vertex two of these strings get contracted due to the presence of the $(C \dots)_{\alpha \beta}$ structure. We can evaluate the remaining contractions of space-time indices using the $d$-dimensional Dirac algebra with an anticommuting $\gamma_5$ and the relation
 \begin{align}
  -C \gamma_\mu C^{-1} &= (\gamma_\mu)^t.
 \end{align}
 This allows us to identify all contributions that are proportional to $\epsilon^0$. Once these are determined we construct the regularized vertices of the irreducible three-quark operators from the linear combinations of the $A$, $V$, $W$ and $U$ operators that were derived at the very beginning. Finally we evaluate the projections of eq. (\ref{eq_ZMSRI_Master}) resulting in the desired scheme matching matrices $Z^{\Gamma,\overline{\text{MS}}\leftarrow \text{mRI}}$.

\subsection{Renormalization Group Behavior}
 Knowing the scaling behavior of $Z$ provides a valuable consistency check of the results. To this end one can compare lattice results that were derived at different renormalization scales with the perturbatively expected scaling. This will be done in subsection \ref{SubS:DataAnalysis}.

 Generally, the scaling behavior of a quantity is described by the renormalization group equation (RGE) and the related beta- and gamma-functions
 \begin{align}
  \beta &= \mu^2 \frac{d}{d\mu^2} \alpha_s(\mu), \\
  \gamma &= -Z^{-1}(\mu) \, \mu^2 \frac{d}{d \mu^2}\, Z(\mu).
  \label{eq_gammabetafunction}
 \end{align}
 Both $\beta$ and the anomalous dimension $\gamma$ can be written as an expansion in the strong coupling:
 \begin{align}
  \frac{\beta(\alpha_s)}{4\pi} &= - \sum\limits_{i=0}^{\infty} \beta_i \left(\frac{\alpha_s(\mu)}{4\pi}\right)^{i+2}, \\
  \gamma(\alpha_s) &= - \sum\limits_{i=0}^{\infty} \frac{\gamma_i}{2} \left(\frac{\alpha_s(\mu)}{4\pi}\right)^{i+1}.
 \end{align}
 In the $\overline{\text{MS}}$ scheme, $\alpha_s$ as well as the beta function are known to high order in perturbative QCD. The gamma function is operator dependent and in most cases not known to the same accuracy.

 Let us focus on the behavior of the renormalization matrix for the three-quark operators under a change of the renormalization scale. It follows from the RGE that a renormalization matrix can be converted from one scale $\mu$ to any other scale $\tilde \mu$ with the help of a scaling matrix $\Delta Z^{\overline{\text{MS}}}$:
 \begin{align}
  \Delta Z^{\overline{\text{MS}}}_{ij}(\mu) Z^{\overline{\text{MS}}}_{jk}(\mu) &= \Delta Z^{\overline{\text{MS}}}_{ij}(\tilde \mu) Z^{\overline{\text{MS}}}_{jk}(\tilde \mu).
  \label{eq_scaleExtrapolationTheory}
 \end{align}
 The matrix $\Delta Z^{\overline{\text{MS}}}(\mu)$ again depends on the beta-function and the operator-specific gamma-function. We derive the scaling matrix for the three-quark operators by independently converting the renormalization matrices of the three-quark vertex and the three quark fields, cf. eq. (\ref{eq_Z}).
 As the behavior of the quark fields is known to higher accuracy, we hope to get a better description for the three-quark operator renormalization by treating this contribution, just as the quark field scheme matching, to order $\alpha_s^2$.

 Hence we apply the expression (\ref{eq_ScalingFunction}) in Appendix \ref{App:ScalingFunction} to convert the wave-function renormalization $Z_q$ together with its anomalous dimension from Appendix \ref{App:QuarkField}. A leading order formula is used for the scaling of the three-quark vertex renormalization matrix $Z^{\Gamma,\overline{\text{MS}}}(\mu)$:
 \begin{align}
  \Delta Z^{\overline{\text{MS}}}_{ij}(\mu) = \left(\alpha_s(\mu)^{-\gamma_0/2\beta_0}\right)_{ij}.
 \end{align}
 We will use this to rescale our results obtained at a set of different $\mu$s to the final scale $\tilde \mu = 2\,\text{GeV}$.

 We still need the gamma function $\gamma^{\Gamma}$ of the three-quark vertex. Omitting contributions that vanish for $\epsilon \to 0$, eq. (\ref{eq_GammaDim}) can be cast in the form
 \begin{align}
  \Gamma^{\text{dim}}_j &= \Gamma_j^{\text{tree}} + \frac{\alpha_s}{4\pi} \tilde \gamma_{jk} \left(\frac{2}{\bar{\epsilon}} 
  - \ln\frac{X^2}{\mu^2} \right) \Gamma_k^{\text{tree}} + \frac{\alpha_s}{4\pi} C_j + \mathcal{O}(\alpha_s^2),
  \label{eq_dim3qVertex}
 \end{align}
 where $X^2$ is some momentum square, $\mu$ the scale of the $\alpha_s$ expansion and $C_j$ is a finite term. We thus find for the renormalization matrix in the $\overline{\text{MS}}$ scheme
 \begin{align}
  Z^{\Gamma,\overline{\text{MS}}}_{ij}(\mu) &= \delta_{ij} - \frac{g_R(\mu)^2}{16\pi^2} \tilde \gamma_{ij} \frac{2}{\bar\epsilon} + \mathcal{O}(g_R^4).
 \end{align}
 Using eq. (\ref{eq_gammabetafunction}) and $g_R(\mu)^2=g^2\mu^{-\epsilon}+\mathcal{O}(g_R^4)$ yields the anomalous dimension matrices:
 \begin{align}
 \gamma^{\Gamma}_{ij} = -\frac{\alpha_s}{4\pi} \tilde\gamma_{ij} + \mathcal{O}(\alpha_s^2).
 \label{eq_gammaGamma}
 \end{align}
 The results for the anomalous dimensions are summarized in Appendix \ref{App:AnomalousDimensions}.

 The whole procedure can only be expected to work in a ``window'', where the renormalization scale is large enough for perturbation theory to be a good approximation and small enough so that the lattice cutoff is sufficiently far away. Hence we expect the condition 
 \begin{align}
  \Lambda_{QCD}^2 \ll \mu^2 \ll \frac{\pi^2}{a^2}
  \label{eq_scalingWindow}
 \end{align}
 to restrict $\mu$ to a reasonable range.

\section{Lattice Calculation and Error Estimation}

\subsection{Details of the Lattice Setup}
 We work with non-perturbatively $\mathcal{O}(a)$ improved clover-Wilson fermions and the plaquette gauge action. We use gauge configurations generated by the QCDSF/UKQCD collaborations with two dynamical flavors of sea quarks, $n_f=2$. The gauge is fixed to Landau gauge by minimizing the functional
 \begin{align}
  F^G[U] = \sum_{x,\mu} \text{Tr} \left( U_\mu^G(x)+ U_\mu^G(x)^\dagger \right) = \sum_{x,\mu} \text{Re Tr}\, U_\mu^G(x),
 \end{align}
 with
 \begin{align}
  U_\mu^G(x) = G(x)\,U_\mu(x)\,G(x+\hat\mu)^\dagger.
 \end{align}
 For any gauge link $U_\mu(x)$ connecting the sites $x$ and $x+\hat\mu$ one can construct gauge equivalent links $U_\mu^G(x)$ by applying $SU(3)$ transformations $G(x)$. Landau gauge is realized by the gauge configuration $U_\mu^G(x)$ that minimizes the functional $F^G[U]$ \cite{Davies:1987vs,Bonnet:1999mj}.

 \begin{table}
  \caption{The used gauge configurations of the QCDSF/UKQCD collaborations. The first column shows the available lattice couplings $\beta$, the second column the lattice volume and the following columns summarizes the hopping parameters $\kappa$.}
  \begin{center}
  \label{tab:lattices}
  \begin{tabular}{ccc}
    $\beta$ & V & $\kappa$   \\
    \hline
    5.20 & $16^3 \times 32$ & 0.13420  \\
    5.20 & $16^3 \times 32$ & 0.13500  \\
    5.20 & $16^3 \times 32$ & 0.13550  \\
    \hline
    5.25 & $16^3 \times 32$ & 0.13460  \\
    5.25 & $16^3 \times 32$ & 0.13520  \\
    5.25 & $24^3 \times 48$ & 0.13575  \\
    5.25 & $24^3 \times 48$ & 0.13600  \\
    \hline
    5.29 & $16^3 \times 32$ & 0.13400  \\
    5.29 & $16^3 \times 32$ & 0.13500  \\
    5.29 & $24^3 \times 48$ & 0.13550  \\
    5.29 & $24^3 \times 48$ & 0.13590  \\
    5.29 & $24^3 \times 48$ & 0.13620  \\
    \hline
    5.40 & $24^3 \times 48$ & 0.13500  \\
    5.40 & $24^3 \times 48$ & 0.13560  \\
    5.40 & $24^3 \times 48$ & 0.13610  \\
    5.40 & $24^3 \times 48$ & 0.13640  \\
  \end{tabular}
 \end{center}
 \end{table}

 \begin{table}
  \caption{Critical hopping parameters $\kappa_c$ and inverse lattice spacings in units of the Sommer parameter $r_0$ extrapolated to the chiral limit.}
  \begin{center}
  \label{tab:latticesChiral}
  \begin{tabular}{cccccccc}
    $\beta$ & $\kappa_\text{c}$ & $r_0/a$  \\
    \hline
    5.20 &  $0.13605$  & $5.454$ \\
    5.25 &  $0.136237$ & $5.880$ \\
    5.29 &  $0.136439$ & $6.201$ \\
    5.40 &  $0.136685$ & $6.946$ \\
  \end{tabular}
  \end{center}
 \end{table}
 We have performed computations on the ensembles of gauge field configurations listed in Table \ref{tab:lattices}. Table \ref{tab:latticesChiral} provides further information on the chiral limit of these lattices. Our approach uses momentum sources \cite{Gockeler:1998ye} which improves the statistics significantly, as it ``averages'' over all space-time coordinates and thus is superior to point-source methods. 
 Due to this benefit an order of ten gauge configurations per ensemble provide sufficiently high statistics for our purposes. Anyhow, the statistical error will turn out to be negligible compared to the systematic error related to the perturbative scheme matching and scale conversion.

 All calculations for the renormalization coefficients were performed on the Regensburg QCDOC machine with partitions of up to $128$ nodes. The code implementation was done in C++ using QDP++, the SciDAC data-parallel programming interface. We have taken over the lattice action and an even-odd-preconditioned stabilized bi-conjugate gradient method for the inversion of the Dirac-operator from the Chroma software library \cite{Edwards:2004sx,Bagel}. 

 For the discretized versions of the first and second covariant derivatives acting on a fermion field inside a three-quark operator we have used
 \begin{align}
  D^x_\mu \psi(x) = \frac{1}{2}& \left( U_\mu(x) \psi(x+\hat\mu) - U^\dagger_\mu(x-\hat\mu)\psi(x-\hat\mu) \right), \\
  D^x_\mu D^x_\nu \psi(x) =\frac{1}{4}& \left( U_\mu(x)U_\nu(x+\hat\mu)\psi(x+\hat\mu+\hat\nu) \right. \nonumber\\
                                    & - U_\mu(x)U_\nu^\dagger(x+\hat\mu-\hat\nu)  \psi(x+\hat\mu-\hat\nu) \nonumber\\
                                    & - U^\dagger_\mu(x-\hat\mu)U_\nu(x-\hat\mu)\psi(x-\hat\mu+\hat\nu) \nonumber\\
                                    & \left. + U^\dagger_\mu(x-\hat\mu)U^\dagger_\nu(x-\hat\mu-\hat\nu)\psi(x-\hat\mu-\hat \nu) \right).
 \end{align}

 To minimize discretization effects, we choose the three external four-momenta $p_i$ of the quarks close to the diagonals of the Brillouin zone. Varying the squares of the momenta allows us to derive the renormalization coefficients at several different values of the renormalization scale $\mu$. The Sommer parameter $r_0=0.467 \,\text{fm}$ \cite{Gockeler:2005rv} is used to set the scale.

\begin{figure}
  \caption{A typical chiral extrapolation for a renormalization coefficient of an operator with two covariant derivatives. The curve shows a linear fit in $1/\kappa$ at a fixed scale $\mu$ in the RI-MOM scheme, the cross marks the chiral limit.}
  \label{Fig:ZChiral}
  \begin{center}
  \includegraphics[scale=0.85,angle=-90]{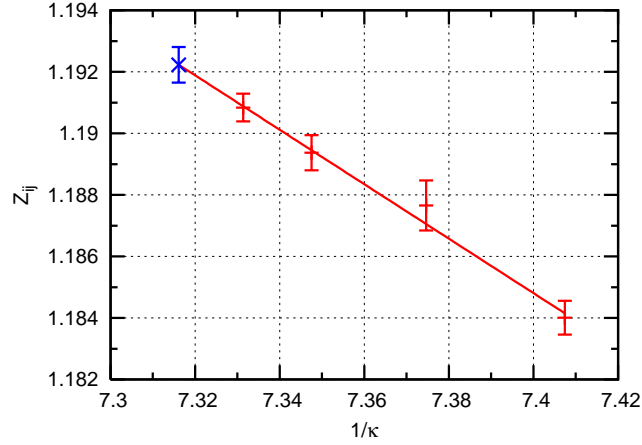}
  \end{center}
 \end{figure}

 At the end we have to perform a chiral extrapolation of the results. Therefore, we compute the amputated three-quark vertices $\Gamma$ for fixed lattice size and fixed coupling $\beta$ at typically three different values of the hopping-parameter $\kappa$. For each $\kappa$ value we then determine the ``renormalization matrix'' $\tilde Z_{ij}^{\text{mRI}}(\mu,\kappa)$. The final mRI renormalization coefficients $Z_{ij}^{\text{mRI}}(\mu)$ are derived from a linear extrapolation in $1/\kappa$ to the critical value $1/\kappa_\text{c}$. A typical chiral extrapolation is depicted in Fig. \ref{Fig:ZChiral} and shows a reasonably linear behavior.

\subsection{Data Analysis}
 \label{SubS:DataAnalysis}
 The Jackknife method is used to estimate the statistical errors for all elements of the matrices $\tilde Z_{ij}^{\text{mRI}}(\mu,\kappa)$. From the uncertainty in the $\chi^2$ fit to the chiral limit we then determine the statistical error for each coefficient of the final result $Z_{ij}^{\text{mRI}}(\mu)$.

 Due to its definition, $Z_{ij}^{\text{mRI}}(\mu)$ may also contain imaginary parts before the scheme matching to $\overline{\text{MS}}$. This can be seen best from eq. (\ref{eq_dim3qVertex}). Applying the projector $P_j$ to the right-hand-side of this equation yields the inverse of the mRI renormalization matrix. It is obvious that, apart from the real terms resulting from the expressions proportional to $\Gamma^{\text{tree}}_i$, we will also end up with a term proportional to $P_j C_i$ from the finite contribution. Written in vector notation this projection reads $\langle v_{P_j} , v_{C_i} \rangle$ and this scalar product is not necessarily real. So finite contributions like $C_i$ may render the mRI renormalization matrix complex. The scheme matching to $\overline{\text{MS}}$ should cancel these terms. However, since we perform a one-loop scheme matching only, we cannot expect a complete cancellation of the imaginary parts. In practice it turns out that the residual imaginary parts are reasonably small and may be used as an independent estimate of the systematic uncertainty induced by the scheme matching. Hence we will quote the real part of the renormalization matrices $Z_{ij}^{\overline{\text{MS}}}(\mu)$ as our results.

 \begin{figure}
  \caption{Scheme matching and scale conversion for a typical diagonal (upper curves) and a typical off-diagonal (lower curves) renormalization coefficient $Z_{ij}$. Dashed curves: mRI$(\mu)$, dash-dotted: $\overline{\text{MS}}(\mu)$, solid: $\overline{\text{MS}}(2\,\text{GeV})$.}
  \label{Fig:ZScaling}
  \begin{center}
  \includegraphics[scale=0.85,angle=-90]{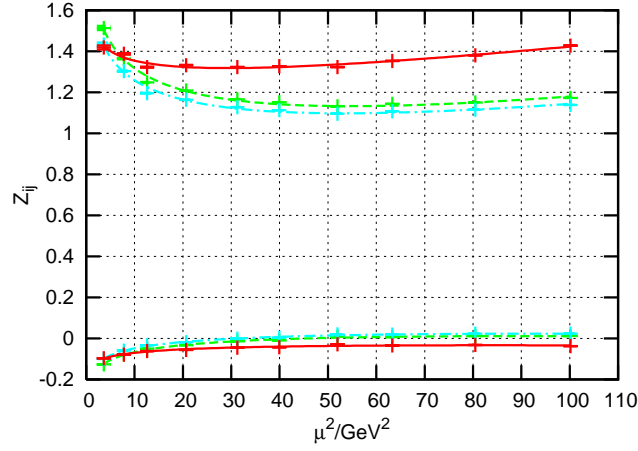}
  \end{center}
 \end{figure}

 \begin{figure}
  \caption{Dependence on the renormalization scale for the same renormalization coefficients as in Fig. \ref{Fig:ZScaling}, however with different external momenta of the three-quark operators. The meaning of the curves is the same as in Fig. \ref{Fig:ZScaling}. One finds good agreement between both momentum geometries.}
  \label{Fig:ZGeometry}
  \begin{center}
  \includegraphics[scale=0.85,angle=-90]{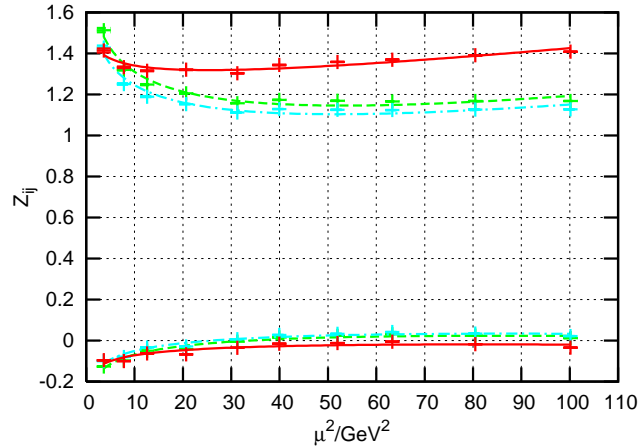}
  \end{center}
 \end{figure}

 The systematic error is estimated via the scaling behavior of the renormalization matrices. To this end we perform a lattice renormalization at ten different renormalization scales $\mu$. We exemplify our procedure with Fig. \ref{Fig:ZScaling}, where we have plotted the scaling of one diagonal and one off-diagonal coefficient of the renormalization matrix for three-quark operators with two derivatives in the irreducible representation $\tau^{\underline{12}}_2$. The general behavior is typical for all representations: While the diagonal coefficient is of order one in all schemes, the mixing off-diagonal coefficient is close to zero. Let us now turn to the different curves in the figure. The dashed curve displays the results in the mRI scheme. As expected it shows a clear dependence on the renormalization scale: In the investigated range $\mu^2 = 3\,\text{GeV}^2 - 100\,\text{GeV}^2$ the renormalization coefficients vary by roughly $40\%$. The scheme matching to $\overline{\text{MS}}$ slightly shifts both mRI curves to lower absolute values (dash-dotted curve). In the third step we convert all results to the same renormalization scale $\tilde \mu=2\,\text{GeV}$ (solid curve). This leads to a manifest flattening of the graph. Ideally the curve would be a constant now, but as already stated neither the lattice, nor the one-loop perturbative approach is free of systematic uncertainties in the whole $\mu^2$ range. At low scales we expect the perturbative expansion to break down, as the coupling becomes too strong for our one-loop approach to hold. This behavior can be observed in the region $\mu^2<10\,\text{GeV}^2$. On the other hand we also have to expect cutoff effects from the lattice calculation at large scales. Also this feature can be found in the figure. In the regime of $\mu^2=10\,\text{GeV}^2\,-\,40\,\text{GeV}^2$ the results appear almost flat, which may be interpreted as a scaling window. Note that although operators with a different number of derivatives may come along with scaling windows shifted to slightly higher or lower values of $\mu^2$, all scaling windows seem to overlap in the quoted region. This indicates that the chosen approach for the non-perturbative renormalization and perturbative scheme matching is well justified.

 Each renormalization coefficient converted to the scale $\tilde \mu=2\,\text{GeV}$ can be reasonably described by the three-parameter fitting formula
 \begin{align}
  Z^{\overline{\text{MS}}}_{ij} = A_{ij} + B_{ij} \log a^2{\mu}^2 + C_{ij} a^2 {\mu}^2,
 \end{align}
 where $\mu$ stands for the original renormalization scale before the conversion.
 This formula incorporates $\mathcal{O}(a^2)$ effects as well as potential further logarithms in $\mu^2$ stemming from higher order corrections in the scheme matching and scale conversion.
Our final result is read off at $\mu^2=20\,\text{GeV}^2$. As our 
estimate of the systematic error we take the maximum of the differences 
with the values at $10\,\text{GeV}^2$ and $40\,\text{GeV}^2$.

 For the two renormalization coefficients plotted in Fig. \ref{Fig:ZScaling} one finds in this way
 \begin{align}
   Z_{11}=1.3370(21)(252), \quad Z_{56} = -0.0528(1)(171),
  \label{eq:typicalZ}
 \end{align}
 where the first error is the statistical and the second the systematic uncertainty.
 As expected for our one-loop calculation, the systematic uncertainty is larger than the statistical error and exceeds it by a typical factor of $\mathcal{O}(10)$ to $\mathcal{O}(100)$. It would be interesting to test our estimates for the systematic error by comparing them with future higher order calculations of the perturbative expressions.

 We have also investigated the dependence of the renormalization matrices on the geometry of the external quark momenta $p_i$. To this end we have calculated the $Z$ matrices for modified angles between the three quark lines running into the vertex. In Fig. \ref{Fig:ZGeometry} we show a typical result for the same renormalization coefficients as in Fig. \ref{Fig:ZScaling}. We find that up to roughly $30\,\text{GeV}^2$ even in the RI-MOM scheme no significant difference between both geometries occurs, whereas at still larger scales small deviations are observed. The scheme-matched curves in the $\overline{\text{MS}}$ scheme lie in most cases on top of each other and the final results are consistent within the systematic errors. For the renormalization coefficients derived with the modified momentum geometry, shown in Fig. \ref{Fig:ZGeometry}, we find
 \begin{align}
  Z_{11}=1.3340(23)(146), \quad Z_{56} = -0.0469(1)(250),
 \end{align} 
 in good agreement with the values in eq. (\ref{eq:typicalZ}).

\section{Results for $Z^{\overline{\text{MS}}}(2\,\text{GeV})$}
Let us now present the renormalization matrices in the 
$\overline{\text{MS}}$ scheme at $\mu=2\,\text{GeV}$ for the 
different $\overline{\text{H}(4)}$ irreducible representations 
$\tau_1^{\underline{4}}$, $\tau_2^{\underline{4}}$, $\tau^{\underline{8}}$,
$\tau_1^{\underline{12}}$, $\tau_2^{\underline{12}}$
of isospin-1/2 three-quark operators. For these representations we use 
the same notation as in Ref.~\cite{Kaltenbrunner:2008pb}. The superscript 
denotes the dimension of the representation, with the underscore 
indicating its spinorial nature. The subscript distinguishes inequivalent 
representations of the same dimension.

 We will denote the statistical error by $E^{\text{st}}$ and the systematic error by $E^{\text{sy}}$. As mentioned, the latter is estimated by comparing the $Z$ values at $20\,\text{GeV}^2$ with those at $10\,\text{GeV}^2$ and $40\,\text{GeV}^2$. Thus the final result reads
 \begin{align}
  Z_{ij} \pm E^{\text{st}}_{ij} \pm E^{\text{sy}}_{ij} .
 \end{align}
 As the statistical errors are almost two orders of magnitude smaller than the systematic errors, we will quote the statistical errors only in two explanatory examples and drop them for the rest of the paper.

Another source of uncertainty results from the error of 
$\Lambda^{\overline{\text{MS}}}$. We use  
$\Lambda^{\overline{\text{MS}}}= 261\,\text{MeV}$ from 
Ref.~\cite{Gockeler:2005rv}. Adding the two given errors we obtain
a combined error of $43\,\text{MeV}$. While the resulting uncertainty is
considerably smaller than $E^{\text{sy}}$ for operators with zero and
one derivative it becomes comparable with $E^{\text{sy}}$ in the case
of two derivatives.

 The mixing multiplets can be read off from Table \ref{tab:1}: Multiplets of the same representation and same dimension can mix under renormalization. Lower-dimensional operators can also mix into higher dimensional ones of the same representation by powers of the inverse lattice spacing $\frac{1}{a}$. We will summarize these $\frac{1}{a}$ and $\frac{1}{a^2}$ admixtures -- wherever present -- in the last columns of the related renormalization matrix (cf. the operator basis in Appendix \ref{App:MSRenormalization}). 

 We find that the results are similar for different values of the lattice coupling $\beta$ and essentially identical at different lattice sizes, which again demonstrates the consistency of the approach. For reasons of better readability, we will only discuss the result for our finest lattice $\beta=5.40$, $L^3\times T=24^3\times48$ in this section, while more details are given in Appendix \ref{App:MSRenormalization}. There we also give the operator bases explicitly.

 \begin{table}[t]
  \caption{Irreducibly transforming multiplets of three-quark operators with isospin 1/2 sorted by their mass dimension. The subscripts $f$, $g$ and $h$ indicate that the covariant derivative(s) act on the first, second or third quark, respectively, cf. \cite{Kaltenbrunner:2008pb}.} 
  \renewcommand{\arraystretch}{1.7}
  \label{tab:1}
  \begin{center}
  \begin{tabular}{|c||c|c|c|}
    \hline
  & dimension 9/2 & dimension 11/2 & dimension 13/2\\
  & (0 derivatives) & (1 derivative) & (2 derivatives) \\
   \hline
   \hline
   $\tau_1^{\underline{4}}$ & $\mathcal{O}_{1}^{(i),\text{MA}}, \, \mathcal{O}_{3}^{(i),\text{MA}}$ & & $\begin{matrix} \mathcal{O}^{(i),\text{MA}}_{ff1}, \, \mathcal{O}^{(i),\text{MA}}_{ff2},\, \mathcal{O}^{(i),\text{MA}}_{ff3}, \\ \mathcal{O}^{(i),\text{MA}}_{gh1}, \, \mathcal{O}^{(i),\text{MA}}_{gh2},\, \mathcal{O}^{(i),\text{MA}}_{gh3} \end{matrix}$\\
   \hline
   $\tau_2^{\underline{4}}$ &  &  & $\begin{matrix} \mathcal{O}^{(i),\text{MA}}_{ff4}, \mathcal{O}^{(i),\text{MA}}_{ff5}, \, \mathcal{O}_{ff6}^{(i),\text{MA}}, \\ \mathcal{O}^{(i),\text{MA}}_{gh4}, \, \mathcal{O}^{(i),\text{MA}}_{gh5}, \, \mathcal{O}_{gh6}^{(i),\text{MA}}\end{matrix}$\\
   \hline
   $\tau^{\underline{8}}$ & &  $\mathcal{O}^{(i),\text{MA}}_{f1}$ & $\begin{matrix} \mathcal{O}^{(i),\text{MA}}_{ff7},\, \mathcal{O}_{ff8}^{(i),\text{MA}}, \, \mathcal{O}^{(i),\text{MA}}_{ff9}, \\ \mathcal{O}^{(i),\text{MA}}_{gh7},\, \mathcal{O}_{gh8}^{(i),\text{MA}}, \, \mathcal{O}^{(i),\text{MA}}_{gh9} \end{matrix}$ \\
   \hline
   $\tau_1^{\underline{12}}$ & $\mathcal{O}_{7}^{(i),\text{MA}}$ & $\begin{matrix} \mathcal{O}^{(i),\text{MA}}_{f2},\\ \mathcal{O}^{(i),\text{MA}}_{f3}, \, \mathcal{O}^{(i),\text{MA}}_{f4} \end{matrix}$  & $\begin{matrix} \mathcal{O}^{(i),\text{MA}}_{ff10},\, \mathcal{O}^{(i),\text{MA}}_{ff11},\, \mathcal{O}^{(i),\text{MA}}_{ff12},\, \mathcal{O}^{(i),\text{MA}}_{ff13},\\ \mathcal{O}^{(i),\text{MA}}_{gh10},\, \mathcal{O}^{(i),\text{MA}}_{gh11},\, \mathcal{O}^{(i),\text{MA}}_{gh12}, \, \mathcal{O}^{(i),\text{MA}}_{gh13} \end{matrix}$ \\
   \hline
   $\tau_2^{\underline{12}}$ &  &  $\begin{matrix} \mathcal{O}^{(i),\text{MA}}_{f5}, \, \mathcal{O}^{(i),\text{MA}}_{f6},\\ \mathcal{O}^{(i),\text{MA}}_{f7}, \, \mathcal{O}^{(i),\text{MA}}_{f8} \end{matrix}$ & $\begin{matrix} \mathcal{O}^{(i),\text{MA}}_{ff14}, \, \mathcal{O}^{(i),\text{MA}}_{ff15}, \, \mathcal{O}^{(i),\text{MA}}_{ff16}, \, \mathcal{O}^{(i),\text{MA}}_{ff17}, \\ \mathcal{O}^{(i),\text{MA}}_{gh14}, \, \mathcal{O}^{(i),\text{MA}}_{gh15}, \, \mathcal{O}^{(i),\text{MA}}_{gh16}, \, \mathcal{O}^{(i),\text{MA}}_{gh17} \end{matrix}$\\
   \hline
  \end{tabular}
  \end{center}
  \renewcommand{\arraystretch}{1.0}
 \end{table}

\subsection{Representation $\tau^{\underline{4}}_1$ (Zero Derivatives)}
 The following renormalization matrix belongs to the three-quark operators without derivatives in the irreducible representation $\tau^{\underline{4}}_1$. There are two multiplets that mix with each other and we find
 \begin{align}
  Z &= \left(
   \begin{array}{cc}
    0.6892 & -0.0285 \\
    -0.0065 & 0.6953
   \end{array}
   \right), \nonumber\\
  E^{\text{st}} &= \left(
   \begin{array}{cc}
    1.7\times 10^{\text{-4}} & 5.4\times 10^{\text{-6}} \\
    2.6\times 10^{\text{-7}} & 1.5\times 10^{\text{-4}}
   \end{array}
   \right), \nonumber\\
  E^{\text{sy}} &= \left(
   \begin{array}{cc}
    0.0151 & 0.0083 \\
    0.0020 & 0.0163
   \end{array}
   \right).
 \end{align}
 The diagonal elements of $Z$ are smaller than one and for both operators of almost equal size. The mixing off-diagonal elements are both negative and amount to roughly four and one per cent of the diagonal coefficients. Furthermore we see that the statistical error is of relative order $10^{-4}$ which renders it negligible compared to the systematic error. The latter one is two orders of magnitude larger and hence will be the dominating source of uncertainty in any renormalization of matrix elements on the lattice.

\subsection{Representation $\tau^{\underline{12}}_1$ (Zero Derivatives)}
 There is only one multiplet of three-quark operators without derivatives belonging to the irreducible representation $\tau^{\underline{12}}_1$. Therefore the renormalization matrix becomes one-dimensional. Our result reads:
 \begin{align}
 Z &= \left(
   \begin{array}{c}
    0.8131
   \end{array}
   \right), \nonumber\\
 E^{\text{sy}} &= \left(
   \begin{array}{c}
    0.0139
   \end{array}
   \right).
 \end{align}
 Again, the diagonal element is smaller than one and the statistical error is of the order $10^{-4}$ so that the systematic error dominates.

\subsection{Representation $\tau^{\underline{8}}$ (One Derivative)}
 We carry on with three-quark operators with one covariant derivative. Since the statistical and systematic errors are of similar order of magnitude as for the above operators without derivatives, we do not quote them here anymore for reasons of better readability, but refer to the summary in Appendix \ref{App:MSRenormalization}. We obtain
 \begin{align}
 Z &= \left(
   \begin{array}{c}
    1.1260
   \end{array}
   \right). \nonumber
 \end{align}
 The diagonal component of the renormalization matrix for operators with one covariant derivative is larger than those for operators without derivatives.

\subsection{Representation $\tau^{\underline{12}}_1$ (One Derivative)}
 For leading-twist operators with one covariant derivative there are three multiplets of the representation $\tau^{\underline{12}}_1$. One lower-dimensional multiplet that belongs to the same irreducible representation and is formed by operators without covariant derivatives can mix with them via one power of the inverse lattice-spacing $\frac{1}{a}$, cf. Table \ref{tab:1} and \cite{Kaltenbrunner:2008pb}. We get
 \begin{align}
  Z &= \left(
   \begin{array}{cccc}
    1.0540 & 0.1081 & -0.0693 & 9.3\times 10^{\text{-4}} \\
    0.0564 & 0.9920 & 0.0483 & -2.0\times 10^{\text{-5}} \\
    0.0033 & -0.0028 & 1.0890 & -2.1\times 10^{\text{-4}}
   \end{array}
   \right). \nonumber
 \end{align}
 The coefficients in the last column, which describe the mixing with lower-dimensional operators, are rather small.

\subsection{Representation $\tau^{\underline{12}}_2$ (One Derivative)}
 This representation has four mixing multiplets. As for $\tau^{\underline{8}}$ there is no mixing with lower-dimensional operators. Therefore these two representations are especially well suited for the evaluation of matrix elements involving three-quark operators with one derivative. We find:
 \begin{align}
  Z &= \left(
   \begin{array}{cccc}
    1.0470 & 0.1066 & -0.0675 & -0.0013 \\
    0.0544 & 0.9898 & 0.0487 & -7.8\times 10^{\text{-4}} \\
    0.0080 & -0.0064 & 1.0870 & 8.4\times 10^{\text{-4}} \\
    -6.4\times 10^{\text{-4}} & -0.0026 & 0.0111 & 1.1320
   \end{array}
   \right). \nonumber
 \end{align}
 The mixing of the second multiplet with the first one is in the realm of ten per cent, while the mixing of the fourth multiplet can almost be neglected. This might be related to the fact that it contains three-quark operators with different chiralities of the quark fields than the other three multiplets.

\subsection{Representation $\tau^{\underline{4}}_1$ (Two Derivatives)}
 We now proceed with leading-twist operators with two covariant derivatives. These operators belong to five inequivalent irreducible representations of the spinorial hypercubic group $\overline{\text{H}(4)}$. We start with the six multiplets of operators with two derivatives in $\tau^{\underline{4}}_1$, which can mix with the two lower-dimensional multiplets, which contain three-quark operators without derivatives. The $\frac{1}{a^2}$-admixture of the latter operators is seen in the last two columns of the renormalization matrix $Z$:
 \begin{scriptsize}
 \begin{align}
  Z &= \left(
   \begin{array}{cccccccc}
    1.3390 & 0.0282 & -0.0010 & 0.0306 & -0.1620 & 0.0693 & 6.6\times \
   10^{\text{-4}} & 6.4\times 10^{\text{-4}} \\
    0.0167 & 1.2950 & -0.0030 & -0.0808 & -0.0458 & 0.0750 & -2.4\times \
   10^{\text{-4}} & -0.0040 \\
    0.0022 & 0.0058 & 1.2710 & -0.0019 & -1.4\times 10^{\text{-4}} & 0.1167 & \
   7.4\times 10^{\text{-5}} & -0.0092 \\
    0.0174 & -0.0892 & 0.0468 & 1.3010 & -0.0803 & 0.0464 & 8.7\times \
   10^{\text{-4}} & 8.3\times 10^{\text{-4}} \\
    -0.0872 & -0.0794 & 0.0708 & -0.0564 & 1.2080 & 0.0615 & 4.0\times \
   10^{\text{-4}} & -0.0020 \\
    0.0249 & 0.0475 & 0.0550 & 0.0285 & 0.0618 & 1.2810 & 2.8\times \
   10^{\text{-5}} & -0.0063
   \end{array}
   \right), \nonumber
 \end{align}
 \end{scriptsize}
$\!\!$Here the statistical errors are by almost a factor of ten larger than for operators without derivatives. However, the relative systematic errors do not seem to change considerably. We furthermore observe a clear hierarchy with the diagonal elements further increasing relative to three-quark operators with one and without derivatives.

\subsection{Representation $\tau^{\underline{4}}_2$ (Two Derivatives)}
 This irreducible representation consists of six leading-twist multiplets. As by group-theoretical arguments mixing with lower-dimensional three-quark operators can be excluded, $\tau^{\underline{4}}_2$ is the representation of choice wherever possible when working with three-quark operators with two derivatives. The renormalization matrix at $2\,\text{GeV}$ reads in the $\overline{\text{MS}}$ scheme:
 \begin{scriptsize}
 \begin{align}
  Z &= \left(
   \begin{array}{cccccc}
    1.2830 & -0.0503 & 0.0142 & 0.0296 & -0.2268 & 0.1382 \\
    -0.0287 & 1.2360 & 0.0137 & -0.0994 & -0.1214 & 0.1365 \\
    -0.0065 & -0.0132 & 1.2820 & 3.8\times 10^{\text{-4}} & -0.0089 & 0.1363 \\
    0.0329 & -0.0138 & 0.0310 & 1.3690 & -0.0637 & -0.0021 \\
    -0.0193 & -0.0231 & 0.0253 & -0.0679 & 1.3330 & 0.0412 \\
    0.0291 & 0.0207 & 0.0301 & 0.0035 & 0.0547 & 1.3770
   \end{array}
   \right).\nonumber
 \end{align}
 \end{scriptsize}

\subsection{Representation $\tau^{\underline{8}}$ (Two Derivatives)}
 Again, six multiplets of operators with two derivatives contribute to this irreducible representation. They can mix with the lower-dimensional operators with one derivative of the representation $\tau^{\underline{8}}$:
 \begin{scriptsize}
 \begin{align}
  Z &= \left(
    \begin{array}{ccccccc}
     1.3150 & 0.0220 & -0.0133 & 0.0139 & 0.1879 & -0.0742 & 0.0067 \\
     -0.0124 & 1.2560 & 0.0090 & 0.0882 & -0.0621 & 0.0653 & -0.0020 \\
     -0.0084 & -0.0176 & 1.2680 & 0.0017 & -0.0123 & 0.1278 & -6.6\times \
    10^{\text{-4}} \\
     0.0349 & 0.0517 & -0.0295 & 1.3240 & 0.0809 & -0.0426 & -0.0053 \\
     0.0691 & -0.0584 & 0.0565 & 0.0543 & 1.2020 & 0.0944 & -1.6\times \
    10^{\text{-4}} \\
     -0.0410 & 0.0549 & 0.0507 & -0.0291 & 0.0715 & 1.2890 & -8.2\times \
    10^{\text{-4}}
    \end{array}
    \right). \nonumber
 \end{align}
 \end{scriptsize}
$\!\!$The mixing coefficients in the last column indicate that the $\frac{1}{a}$-contribution of the lower-dimensional multiplet might be important in practical applications.

 \subsection{Representation $\tau^{\underline{12}}_1$ (Two Derivatives)}
 Eight operator multiplets with two derivatives mix with the three multiplets with one derivative and the operator multiplet without derivatives of $\tau^{\underline{12}}_1$. For reasons of better readability we split off the last four columns of the renormalization matrix, which describe the mixing with these lower-dimensional operators, and summarize the coefficients in a separate matrix $Z'$. Then the $\frac{1}{a}$ mixing with the operators with one derivatives can be read off from the first three columns of $Z'$, the $\frac{1}{a^2}$ mixing from its last column:
 \begin{scriptsize}
 \begin{align}
  Z &= \left(
   \begin{array}{cccccccccccc}
    1.3020 & -0.0515 & 0.0317 & 0.0039 & 0.0184 & -0.1943 & 0.0940 & -8.9\times \
   10^{\text{-4}}  \\
    -0.0212 & 1.2450 & 0.0210 & -4.3\times 10^{\text{-4}} & -0.1064 & -0.1106 & \
   0.1174 & 7.9\times 10^{\text{-4}}  \\
    0.0112 & 0.0113 & 1.2820 & -5.7\times 10^{\text{-4}} & 0.0186 & 0.0579 & \
   0.0845 & -4.4\times 10^{\text{-4}} \\
    0.0013 & -0.0046 & 0.0019 & 1.3220 & -0.0093 & -0.0166 & 0.0071 & 0.1027 \\
    0.0749 & -0.0050 & 0.0329 & 7.6\times 10^{\text{-4}} & 1.3470 & -0.0019 & \
   -0.0359 & -3.2\times 10^{\text{-4}}  \\
    -0.0335 & -0.0137 & 0.0328 & 1.0\times 10^{\text{-4}} & -0.0483 & 1.3000 & \
   0.0382 & -9.3\times 10^{\text{-4}}  \\
    0.0617 & 0.0676 & 0.0472 & 0.0010 & 0.0182 & 0.1031 & 1.2900 & 1.7\times \
   10^{\text{-4}}  \\
    -7.9\times 10^{\text{-4}} & 0.0089 & -0.0043 & 0.0158 & 0.0065 & 0.0027 & \
   -0.0024 & 1.3560 
   \end{array}
   \right), \nonumber\\
  Z' &= \left(
   \begin{array}{cccccccccccc}
    -9.4\times 10^{\text{-4}} & -4.5\times 10^{\text{-4}} & \
   0.0046 & 0.0208 \\
     -0.0059 & 0.0060 & -0.0051 & 0.0139 \\
     -0.0017 & 0.0052 & -0.0033 & 0.0025 \\
      0.0057 & -0.0188 & 0.0091 & -3.0\times 10^{\text{-4}} \\
      0.0020 & -0.0041 & 0.0070 & -0.0236 \\
     -0.0044 & 0.0066 & -0.0084 & -0.0200 \\
     -0.0046 & 0.0029 & -0.0075 & -0.0083 \\
     0.0101 & -0.0046 & -1.8\times 10^{\text{-4}} & -9.4\times \
   10^{\text{-4}}
   \end{array}
   \right). \nonumber
 \end{align}
 \end{scriptsize}
$\!\!$Compared to the diagonal elements around $1.3$, the sixth operator mixes into the first with $0.2$ and the $\frac{1}{a^2}$ mixing contributes with a factor of up to $0.02$. When renormalizing matrix elements with these coefficients, the mixing of the lower-dimensional lattice operators can only be neglected, if the matrix elements of these operators are considerably smaller than those containing operators with two derivatives. Whether this is the case or not must be decided in each individual case. As already stated earlier, whenever possible one should anyway make use of the operators of representation $\tau^{\underline{4}}_2$, because these do not suffer from mixing with lower-dimensional operators.

 \subsection{Representation $\tau^{\underline{12}}_2$ (Two Derivatives)}
 This last representation also contains eight three-quark operators with two derivatives that can mix with the four lower-dimensional operators with one derivative of the same representation. For reasons of better readability we have again split off the last four columns of the renormalization matrix and display the related coefficients in the separate matrix $Z'$: 
 \begin{scriptsize}
 \begin{align}
  Z &= \left(
   \begin{array}{cccccccccccc}
    1.3340 & -0.0073 & 0.0235 & -0.0019 & -0.0065 & -0.2075 & 0.0934 & 0.0061 \\
    -0.0048 & 1.2800 & -5.5\times 10^{\text{-4}} & -1.1\times 10^{\text{-4}} & \
   -0.0925 & -0.0853 & 0.0902 & -3.2\times 10^{\text{-4}}  \\
    0.0124 & 0.0037 & 1.2850 & -1.2\times 10^{\text{-4}} & -0.0060 & -0.0017 & \
   0.1101 & -4.2\times 10^{\text{-5}}  \\
    0.0059 & 0.0091 & -0.0024 & 1.3290 & 0.0014 & 0.0078 & -0.0142 & 0.1231  \\
    0.0301 & -0.0645 & 0.0426 & 8.1\times 10^{\text{-5}} & 1.3430 & -0.0877 & \
   0.0288 & 0.0013  \\
    -0.0581 & -0.0559 & 0.0530 & 2.5\times 10^{\text{-4}} & -0.0606 & 1.2850 & \
   0.0649 & -0.0015  \\
    0.0298 & 0.0263 & 0.0589 & -2.7\times 10^{\text{-4}} & -0.0027 & 0.0605 & \
   1.3380 & -3.3\times 10^{\text{-4}} \\
    0.0047 & -4.3\times 10^{\text{-4}} & 9.7\times 10^{\text{-5}} & 0.0401 & \
   -0.0015 & 0.0043 & -0.0015 & 1.3740 
   \end{array}
   \right), \nonumber\\
  Z' &= \left(
   \begin{array}{cccccccccccc}
    -0.0017 & -0.0021 & 5.5\times 10^{\text{-4}} & -0.0026 \\
     0.0116 & -0.0152 & 0.0114 & 1.1\times 10^{\text{-4}} \\
     0.0025 & -4.2\times 10^{\text{-5}} & 0.0034 & -9.0\times 10^{\text{-6}} \\
     0.0028 & -1.1\times 10^{\text{-4}} & 0.0068 & -7.7\times 10^{\text{-5}} \\
    -0.0089 & 0.0085 & -0.0066 & 0.0040 \\
     0.0074 & -0.0057 & 0.0037 & -1.0\times 10^{\text{-4}} \\
     0.0083 & -0.0028 & 0.0025 & 0.0021 \\
    -0.0158 & 0.0321 & 0.0059 & -1.3\times 10^{\text{-5}}
   \end{array}
   \right). \nonumber
 \end{align}
 \end{scriptsize}
$\!\!$Also in this case the renormalization matrix has mixing coefficients of a few per cent for the lower-dimensional operators. Whether or not they have to be taken into account depends again on the magnitude of the corresponding matrix elements.

\section{Renormalization of Moments of the Nucleon Distribution Amplitude}
 Having presented the renormalization matrices in the previous section, the next step is their application to the moments of the nucleon distribution amplitude $\phi$:
 \begin{align}
  \phi^{lmn}(\mu) &= \int [dx_i] \,\,\, x_1^l x_2^m x_3^n \, \phi(x_1,x_2,x_3,\mu), \label{eq:DAMoments}\\
  [dx_i] &= dx_1 dx_2 dx_3 \delta(1-x_1-x_2-x_3). \label{eq:DAIntegrationMeasure}
 \end{align}
 For a detailed discussion of the calculation of these quantities in lattice QCD, the notation used and the physical interpretation of the results we refer to \cite{Gockeler:2008xv,Niko_Forthcoming}. Here we focus on the behavior under renormalization and quote numbers only to exemplify the involved orders of magnitude.

\subsection{Zeroth Moment}
 The zeroth moment is linked to a matrix element of three-quark operators without derivatives that is proportional to $f_N$, the normalization constant of the nucleon wave function:
 \begin{align}
  \langle 0 \vert \mathcal{O}^{000}_{A,0} \vert p \rangle &= f_N (ip_1\gamma_1 -ip_2\gamma_2) N(p).
 \end{align}
 Here $N$ denotes the nucleon spinor. The definition of $\mathcal{O}^{000}_{A,0}$ can be found in \cite{Niko_Forthcoming}. For the renormalization of the matrix element only the multiplet and the irreducible representation it belongs to is relevant. The first spinor component reads in our MA-isospin operator-basis:
 \begin{align}
  \left(\mathcal{O}^{000}_{A,0}\right)_1 &= - \frac{4\sqrt{2}}{3} \, \mathcal{O}_{7}^{(6),\text{MA}}.
 \end{align}
 This operator belongs to the irreducible representation $\tau^{\underline{12}}_1$. Hence the bare value of $f_N$ is renormalized by multiplication with $Z(\tau^{\underline{12}}_1)$:
 \begin{align}
  f_N^{\overline{\text{MS}}} &= Z(\tau^{\underline{12}}_1) \, f_N.
 \end{align}
 With $f_N/m_N^2 \approx 4.3 \cdot 10^{-3}$ we find $f_N^{\overline{\text{MS}}}/m_N^2 \approx 3.5 \cdot 10^{-3}$, cf. Table \ref{Tab:RenormalizedMoments}.

\subsection{Next-to-Leading Twist Constants $\lambda_1$, $\lambda_2$}
 For three-quark operators without derivatives we have also computed the renormalization matrix for next-to-leading twist. This enables us to renormalize the constants $\lambda_1$ and $\lambda_2$, which describe the coupling of the nucleon to two different interpolating fields used in QCD sum rules. Since we work in Euclidean space we have:
 \begin{align}
  \langle 0 \vert \mathcal{L}_\tau(0) \vert p \rangle &= - \lambda_1 m_N N_\tau (p), \nonumber\\
  \langle 0 \vert \mathcal{M}_\tau(0) \vert p \rangle &= - \lambda_2 m_N N_\tau (p).
 \end{align}
 We perform the following discussion in detail to clarify the treatment of mixing matrix elements and the required change of basis.
 In a first step we relate the operators sandwiched between the nucleon and vacuum to our MA-isospin basis:
 \begin{align}
  \mathcal{L}_\tau &= -8\, \mathcal{O}_{3}^{(\tau),\text{MA}}, \nonumber\\
  \mathcal{M}_\tau &= \frac{16}{\sqrt{3}}\, \mathcal{O}_{1}^{(\tau),\text{MA}}.
 \end{align}
 Hence $\lambda_1$ renormalizes like $-8\, \mathcal{O}_{3}^{(\tau),\text{MA}}$ and $\lambda_2$ like $\frac{16}{\sqrt{3}}\, \mathcal{O}_{1}^{(\tau),\text{MA}}$. The operators $\mathcal{O}_{1}^{(\tau),\text{MA}}$ and $\mathcal{O}_{3}^{(\tau),\text{MA}}$ form the basis for our renormalization matrix for the representation $\tau^{\underline{4}}_1$ and mix with each other under renormalization, i.e., introducing the abbreviations
 \begin{align}
  L_1 = \frac{\sqrt{3}}{16}\, \lambda_2, \nonumber\\
  L_2 = -\frac{1}{8}\, \lambda_1,
 \end{align}
 we have
 \begin{align}
  L_1^{\text{ren}} = Z_{1i}(\tau_1^{\underline{4}})\, L_i, \nonumber\\
  L_2^{\text{ren}} = Z_{2i}(\tau_1^{\underline{4}})\, L_i.
 \end{align}
 Substituting back the bare $\lambda$s yields finally the desired relation between the bare and renormalized values:
 \begin{align}
  \lambda_1^{\text{ren}} &= -Z(\tau_1^{\underline{4}})_{21} \,\frac{\sqrt{3}}{2}\,\lambda_2 + Z(\tau_1^{\underline{4}})_{22}\,\lambda_1, \nonumber\\
  \lambda_2^{\text{ren}} &= Z(\tau_1^{\underline{4}})_{11} \,\lambda_2 - \frac{2}{\sqrt{3}}\, Z(\tau_1^{\underline{4}})_{12}\,\lambda_1.
 \end{align}

\begin{table}
 \caption{Bare and renormalized values for (combinations of) the lowest moments of the nucleon distribution amplitude. In the case of the renormalized values the first error is statistical, while the second error estimates the systematic uncertainties due to renormalization and chiral extrapolation. 
The data are taken from \cite{Niko_Forthcoming}.}
 \label{Tab:RenormalizedMoments}
 \begin{center}
 \begin{tabular}{c|ccc}
       & $f_N/m_N^2$             & $\lambda_1$/GeV$^2$          & $\lambda_2$/GeV$^2$ \\
  \hline
  bare & 0.00429(7) & -0.0729(14) & 0.1464(27)   \\
  ren. & 0.00349(6)(12) & -0.0498(9)(42) & 0.0985(19)(87)\\
  \hline
  \hline
  & $\phi^{100}$ & $\phi^{010}$ & $\phi^{001}$ \\
  \hline
  bare & 0.294(6) & 0.272(6) & 0.274(6)              \\
  ren. & 0.346(8)(9) & 0.312(8)(13) & 0.314(8)(10)\\
  \hline
  \hline
  & $\phi^{200}$ & $\phi^{020}$ & $\phi^{002}$ \\
  \hline
  bare & 0.113(7) & 0.095(6) & 0.106(6)\\
  ren. & 0.152(11)(83) & 0.127(10)(25) & 0.140(10)(17)\\
  \hline
  \hline
  & $\phi^{011}$ & $\phi^{101}$ & $\phi^{110}$ \\
  \hline
  bare & 0.065(4) & 0.069(6) & 0.071(4) \\
  ren. & 0.084(7)(31) & 0.112(9)(31) & 0.105(7)(2)\\
  \hline
  \hline
       & $S_1$     & $S_2$          \\
  \hline
  bare & 0.840(10) & 0.722(16) \\
  ren. & 0.972(12)(13) & 1.021(28)(98)\\
  \hline
  \hline
       & $S^{100}$ & $S^{010}$ & $S^{001}$ \\
  \hline
  bare & 0.252(10) &  0.230(8) & 0.240(9) \\
  ren. & 0.370(14)(63) & 0.316(13)(16) & 0.336(14)(20)
 \end{tabular}
 \end{center}
\end{table}

\subsection{First Moments}
 The first moments of the nucleon distribution amplitude can be derived from matrix elements of three-quark operators with one covariant derivative:
 \begin{align}
  \langle 0 \vert \mathcal{O}^{lmn}_{A,1} \vert p \rangle = f_N \phi^{lmn} [(p_1\gamma_1 -p_2\gamma_2)(ip_3\gamma_3-E(\vec{p})\gamma_4) -2ip_1p_2\gamma_1\gamma_2] N(p).
  \label{Eq:Phi1}
 \end{align}
 The operators $\mathcal{O}^{lmn}_{A,1}$, where $l+m+n=1$, are related to the MA operators of the representation $\tau^{D\underline{12}}_2$:
 \begin{align}
  \left(\mathcal{O}^{100}_{A,1}\right)_1 &= \frac{8}{3} \, (\mathcal{O}_{f6}^{(1),\text{MA}} - \mathcal{O}_{f7}^{(1),\text{MA}}),\nonumber\\
  \left(\mathcal{O}^{010}_{A,1}\right)_1 &= \frac{8}{3} \, (- \mathcal{O}_{f5}^{(1),\text{MA}}),\nonumber\\
  \left(\mathcal{O}^{001}_{A,1}\right)_1 &= \frac{8}{3} \, \mathcal{O}_{f6}^{(1),\text{MA}}. 
  \end{align}
 The matrix elements of these operators yield the bare values for $f_N \phi^{100}$, $f_N \phi^{010}$ and $f_N \phi^{001}$, as eq. (\ref{Eq:Phi1}) shows. Using the definitions
 \begin{align}
  M_1 &= -\phi^{010}, \nonumber\\
  M_2 &=  \phi^{001}, \nonumber\\
  M_3 &=  \phi^{001} - \phi^{100},
 \end{align}
 we obtain the following relation to the renormalized first moments of the nucleon distribution amplitude:
 \begin{align}
  f_N^\text{ren}\, \phi^{100,\text{ren}} &= \left(Z(\tau^{D\underline{12}}_2)_{2i} - Z(\tau^{D\underline{12}}_2)_{3i}\right) \,f_N M_i, \nonumber\\
  f_N^\text{ren}\, \phi^{010,\text{ren}} &= -Z(\tau^{D\underline{12}}_2)_{1i}\, f_N M_i, \nonumber\\
  f_N^\text{ren}\, \phi^{001,\text{ren}} &=  Z(\tau^{D\underline{12}}_2)_{2i}\, f_N M_i.
 \end{align}
 Upon using $f_N^\text{ren} = Z(\tau^{\underline{12}}_1) \, f_N$ we get:
 \begin{align}
  \phi^{100,\text{ren}} &= \frac{1}{Z(\tau^{\underline{12}}_1)}\, \left(Z(\tau^{D\underline{12}}_2)_{2i} - Z(\tau^{D\underline{12}}_2)_{3i}\right) M_i, \nonumber\\
  \phi^{010,\text{ren}} &= -\frac{1}{Z(\tau^{\underline{12}}_1)}\, Z(\tau^{D\underline{12}}_2)_{1i} M_i, \nonumber\\
  \phi^{001,\text{ren}} &=  \frac{1}{Z(\tau^{\underline{12}}_1)}\, Z(\tau^{D\underline{12}}_2)_{2i} M_i.
 \end{align}
 Due to eqs. (\ref{eq:DAMoments}) and (\ref{eq:DAIntegrationMeasure}) the moments must comply with
 \begin{align}
  S_1 := \phi^{100}+ \phi^{010}+ \phi^{001} &\stackrel{!}{=} \phi^{000} \equiv 1
 \end{align}
 While the bare values do not fulfill this equation (the sum equals $0.84$), the sum of the renormalized values is found to be $0.97$. Within errors this is in agreement with the constraint.

\subsection{Second Moments}
 Analogously we can renormalize the second moments of the nucleon distribution amplitude. A typical matrix element is given by
 \begin{align}
  \langle 0 \vert \mathcal{O}^{lmn}_{2} \vert p \rangle = f_N \phi^{lmn}\,& [p_1p_2\gamma_1\gamma_2(ip_3\gamma_3 +E(\vec{p})\gamma_4)+ \nonumber\\ 
  &+ ip_3E(\vec{p})\gamma_3\gamma_4(ip_1\gamma_1-ip_2\gamma_2)] N(p).
  \label{Eq:Phi2}
 \end{align}
 Again we investigate the multiplets and the representation to which the operators $\mathcal{O}^{lmn}_{2}$, $l+m+n=2$, belong by rewriting them in terms of our isospin mixed-antisymmetric basis. For the fourth spinor component we have, e.g.,
 \begin{align}
  \left( \mathcal{O}^{200}_{2}\right )_4 &= \frac{4\sqrt{2}}{3\sqrt{3}} \, \left( \mathcal{O}_{ff5}^{(1),\text{MA}} - \mathcal{O}_{ff6}^{(1),\text{MA}} \right), \nonumber\\
  \left( \mathcal{O}^{020}_{2}\right )_4 &= \frac{4\sqrt{2}}{3\sqrt{3}} \, \mathcal{O}_{ff4}^{(1),\text{MA}}, \nonumber\\
  \left( \mathcal{O}^{002}_{2}\right )_4 &= \frac{4\sqrt{2}}{3\sqrt{3}} \, \mathcal{O}_{ff5}^{(1),\text{MA}}, \nonumber\\
  \left( \mathcal{O}^{011}_{2}\right )_4 &= \frac{4\sqrt{2}}{3\sqrt{3}} \, \left( \mathcal{O}_{gh5}^{(1),\text{MA}} - \mathcal{O}_{gh6}^{(1),\text{MA}} \right), \nonumber\\
  \left( \mathcal{O}^{101}_{2}\right )_4 &= \frac{4\sqrt{2}}{3\sqrt{3}} \, \mathcal{O}_{gh4}^{(1),\text{MA}}, \nonumber\\
  \left( \mathcal{O}^{110}_{2}\right )_4 &= \frac{4\sqrt{2}}{3\sqrt{3}} \, \mathcal{O}_{gh5}^{(1),\text{MA}}.
 \end{align}
 With the new observables
 \begin{align}
  M_1' &= \phi^{020}, \nonumber\\
  M_2' &= \phi^{002}, \nonumber\\
  M_3' &= \phi^{002} - \phi^{200}, \nonumber\\
  M_4' &= \phi^{101}, \nonumber\\
  M_5' &= \phi^{110}, \nonumber\\
  M_6' &= \phi^{110} - \phi^{011}
 \end{align}
 we find the renormalized quantities
 \begin{align}
  \phi^{200,\text{ren}} &= \frac{1}{Z(\tau^{\underline{12}}_1)}\,\left(Z(\tau_2^{DD\underline{4}})_{2i} - Z(\tau_2^{DD\underline{4}})_{3i}\right) M_i',  \nonumber\\
  \phi^{020,\text{ren}} &= \frac{1}{Z(\tau^{\underline{12}}_1)}\,Z(\tau_2^{DD\underline{4}})_{1i} M_i', \nonumber\\
  \phi^{002,\text{ren}} &= \frac{1}{Z(\tau^{\underline{12}}_1)}\,Z(\tau_2^{DD\underline{4}})_{2i} M_i', \nonumber\\
  \phi^{011,\text{ren}} &= \frac{1}{Z(\tau^{\underline{12}}_1)}\,\left(Z(\tau_2^{DD\underline{4}})_{5i} - Z(\tau_2^{DD\underline{4}})_{6i}\right) M_i', \nonumber\\
  \phi^{101,\text{ren}} &= \frac{1}{Z(\tau^{\underline{12}}_1)}\,Z(\tau_2^{DD\underline{4}})_{4i} M_i', \nonumber\\
  \phi^{110,\text{ren}} &= \frac{1}{Z(\tau^{\underline{12}}_1)}\,Z(\tau_2^{DD\underline{4}})_{5i} M_i'.
 \end{align}
 Here, the renormalized values must fulfill four constraints:
 \begin{align}
 S^{100} = \phi^{200}+ \phi^{110} +\phi^{101} &\stackrel{!}{=} \phi^{100}, \nonumber\\
 S^{010} = \phi^{020}+ \phi^{110} +\phi^{011} &\stackrel{!}{=} \phi^{010}, \nonumber\\
 S^{001} = \phi^{002}+ \phi^{101} +\phi^{011} &\stackrel{!}{=} \phi^{001}, \nonumber\\
 S_2 := S^{100} + S^{010} + S^{001} &\stackrel{!}{=} 1.
 \end{align}
 Comparing the results in Table \ref{Tab:RenormalizedMoments} reveals good agreement within errors also for these quantities. This is an encouraging result, which demonstrates the consistency of the applied renormalization. It also makes us confident that we did not severely underestimate our errors.

\section{Summary and Conclusions}
 In this paper we have set up a lattice renormalization scheme for three-quark operators based on the RI-MOM approach. The introduction of a suitable set of projectors facilitates the definition and practical evaluation of a renormalization matrix in our mRI scheme. In a second step we have performed a scheme matching to $\overline{\text{MS}}$ and a scale conversion to $2\,\text{GeV}$ based on one-loop continuum perturbation theory. It was found that the statistical errors are negligible compared to the systematic uncertainties. Finally we have explained how to apply our results to the renormalization of low moments of the nucleon distribution amplitude.

\section{Acknowledgements}
 The numerical simulations have been performed on the Hitachi SR8000 at LRZ (Munich), apeNEXT and APEmille at NIC/DESY (Zeuthen) and BlueGene/Ls at NIC/JSC (J\"ulich), EPCC (Edinburgh) as well as QCDOC (Regensburg) using the Chroma software library \cite{Edwards:2004sx,Bagel}. This work was supported by DFG (Forschergruppe ``Gitter-Ha\-dro\-nen-Ph\"anomenologie'' and SFB/TR 55 ``Hadron Physics from Lattice QCD''), by EU I3HP (contract No. RII3-CT-2004-506078) and by BMBF. TK wants to thank J. Bloch for helpful discussions.

\appendix
\section{Perturbation Theory for the Three-Quark Operators}
 \label{App:FeynmanDiagrams}

\subsection{The Action}
 \label{App:PTAction}
 In this appendix we summarize the conventions that we have adopted for the perturbative evaluation of the three-quark vertex $\Gamma$ in one-loop order. We use the Euclidean action
 \newcommand{\s}[1]{#1\!\!\!/}
 \begin{align}
  S_E = \int d^4x [&\bar \psi (\gamma_\mu D_\mu+m)\psi + \frac{1}{4}F_{\mu\nu}^a F_{\mu\nu}^a + \frac{1}{2\xi} (\partial_\mu A_\mu^a)  (\partial_\nu A_\nu^a)\nonumber\\
  &-\bar u^a \partial_\mu (\partial_\mu \delta_{ab} - g f_{abc}A_\mu^c)u^b ],
 \end{align}
 with
 \begin{align}
  D_\mu &= \partial_\mu-ig\lambda_a A_\mu^a, \nonumber\\
  F_{\mu\nu}^a &= \partial_\mu A^a_\nu-\partial_\nu A_\mu^a + g f_{abc}A_\mu^b A_\nu^c. 
 \end{align}
 Here $\psi$ denotes the fermion fields, $u$ the ghosts, $A_\mu^a$ the gluon field and $\lambda_a$ stands for the Gell-Mann matrices fulfilling $[\lambda_a, \lambda_b]=if_{abc}\lambda_c$.
 The propagators in momentum space are given by:
 \begin{align}
  S(p)_{\alpha \beta} &= \frac{(-i \s{p}+m)_{\alpha\beta}}{p^2+m^2}, \nonumber\\
  G(p)_{\mu \nu} &= \frac{1}{p^2}\left(\text{g}_{\mu\nu}-(1-\xi)\frac{p_\mu p_\nu}{p^2}\right).
 \end{align}
 We work in the chiral limit $m=0$ and use the Landau gauge $\xi=0$.

\subsection{The Scaling Function}
 \label{App:ScalingFunction}
 A three-loop formula for the function $\Delta Z^{\overline{MS}}(\mu)$ is given by:
 \begin{align}
  \Delta Z^{\overline{MS}}(\mu)^{-1} = \alpha_s(\mu)^{\bar \gamma_0} \Big(& 1+ \frac{\alpha_s(\mu)}{4\pi} (\bar \gamma_1 - \bar \beta_1 \bar \gamma_0 ) +\frac{1}{2} \left( \frac{\alpha_s(\mu)}{4\pi} \right)^2 \big( (\bar \gamma_1 - \bar \beta_1 \bar \gamma_0 )^2 \nonumber\\
 & + \bar \gamma_2 + \bar \beta_1^2 \bar \gamma_0 - \bar \beta_1 \bar \gamma_1 - \bar \beta_2 \bar \gamma_0 \big) \Big) +\mathcal{O}(\alpha_s^3), 
  \label{eq_ScalingFunction}
 \end{align}
 with the abbreviations $\bar \beta_i = \beta_i / \beta_0$ and $\bar \gamma_i = \gamma_i / (2\beta_0)$.

\subsection{The Quark Field Anomalous Dimension}
 \label{App:QuarkField}
 Finally we give the anomalous dimension of the quark field so that we can extract the scaling behavior from eq. (\ref{eq_ScalingFunction}). It reads \cite{Chetyrkin:2000dq}:
 \begin{align}
  \gamma_q =& +\frac{\alpha_s}{4\pi}\, C_F (-\xi) \nonumber\\
  &+\frac{\alpha_s^2}{(4\pi)^2}\, C_A\, C_F (-\frac{25}{4} -2\,\xi -\frac{1}{4}\,\xi^2) +\frac{\alpha_s^2}{(4\pi)^2}\,2\,C_F\, n_f\, T + \frac{\alpha_s^2}{(4\pi)^2}\, \frac{3}{2}\, C_F^2 \nonumber\\
  &+\frac{\alpha_s^3}{(4\pi)^3}\, C_A^2\, C_F ( -\frac{9155}{144} -\frac{3}{4}\, \zeta_3\, \xi - \frac{3}{8}\, \zeta_3\, \xi^2 +\frac{69}{8}\, \zeta_3 - \frac{263}{32}\, \xi - \frac{39}{32}\, \xi^2 -\frac{5}{16}\, \xi^3 ) \nonumber\\
  &+\frac{\alpha_s^3}{(4\pi)^3}\, C_A\, C_F\, n_f\, T (\frac{287}{9} +\frac{17}{4}\, \xi) +\frac{\alpha_s^3}{(4\pi)^3}\, C_A\, C_F^2\, (\frac{143}{4} -12\, \zeta_3 )\nonumber\\
  &+\frac{\alpha_s^3}{(4\pi)^3}\, C_F\, n_f^2\, T^2 (\frac{-20}{9}) + \frac{\alpha_s^3}{(4\pi)^3} (-3) C_F^2\, n_f\, T + \frac{\alpha_s^3}{(4\pi)^3}\, C_F^3 (-\frac{3}{2}).\nonumber
 \end{align}
 Here
 \begin{align}
  C_A = 3, \quad C_F = \frac{4}{3}, \quad T=\frac{1}{2}.
 \end{align}

\section{Anomalous Dimensions}
 \label{App:AnomalousDimensions}
 In the following we summarize the gauge invariant anomalous dimensions $\gamma$ for the three-quark operators. In the one-loop perturbative expansion they are identical to the anomalous dimension $\gamma^\Gamma$, eq. (\ref{eq_gammaGamma}), of the three-quark vertex in Landau gauge:
 \begin{align}
  \gamma = \gamma^\Gamma\vert_{\xi=0} + \mathcal{O}(\alpha_s^2).
 \end{align}
 The operator basis is taken over from the linearly independent, isospin-1/2 $\overline{\text{H}(4)}$ irreducibly transforming multiplets of three-quark operators introduced in Appendix B of \cite{Kaltenbrunner:2008pb}. We have checked that the eigenvalues of our anomalous dimension matrices agree with those presented by Peskin in \cite{Peskin:1979mn}.

\subsection{Operators without Derivatives}
 We start with three-quark operators without covariant derivatives. For the irreducible representation $\tau^{\underline 4}_1$ we choose the operator basis
 \begin{align}
  O_1 = \mathcal{O}_1^{(i),\text{MA}}, &\quad O_2 = \mathcal{O}_3^{(i),\text{MA}}. \nonumber
 \end{align}
 To first order we find the anomalous dimension matrix
 \begin{align}
  \gamma^\Gamma = \frac{\alpha_s}{4\pi} \left(
  \begin{array}{cc}
   2 (\xi +1) & 0 \\
   0 & 2 (\xi +1)
  \end{array}
  \right).
 \end{align}
 The operator basis for the representation $\tau^{\underline 12}_1$ is taken to be $O_1= \mathcal{O}_7^{(i),\text{MA}}$. Then
 \begin{align}
  \gamma^\Gamma = \frac{\alpha_s}{4\pi} \left(
  \begin{array}{c}
   2\,\xi -\frac{2}{3}
  \end{array}
  \right).
 \end{align}

\subsection{Operators with One Derivative}
 In the following we summarize the anomalous dimensions for three-quark operators with one covariant derivative. The representation $\tau^{\underline 8}$ has the operator basis $O_1= \mathcal{O}^{(i),\text{MA}}_{f1}$ with
 \begin{align}
  \gamma^\Gamma = \frac{\alpha_s}{4\pi} \left(
  \begin{array}{c}
   2\,\xi -4
  \end{array}
  \right).
 \end{align}
 For the irreducible representation $\tau^{D\underline{12}}_1$ we choose the operator basis
 \begin{alignat}{3}
  O_1 &= \mathcal{O}_{f2}^{(i),\text{MA}}, \quad & O_2 &= \mathcal{O}_{f3}^{(i),\text{MA}}, \quad &
  O_3 &= \mathcal{O}_{f4}^{(i),\text{MA}}. \nonumber
 \end{alignat}
 Then the anomalous dimension is given by
 \begin{align}
  \gamma^\Gamma = \frac{\alpha_s}{4\pi} \left(
  \begin{array}{ccc}
   2\,\xi -\frac{22}{9}\, & -\frac{16}{9} & \frac{8}{9} \\[2mm]
   -\frac{8}{9} & 2\,\xi -\frac{14}{9}\, & -\frac{2}{3} \\[2mm]
   0 & 0 & 2\,\xi -\frac{26}{9}\,
  \end{array}
  \right).
 \end{align}
 In the basis 
 \begin{alignat}{4}
  O_1 &= \mathcal{O}_{f5}^{(i),\text{MA}}, \quad & O_2 &= \mathcal{O}_{f6}^{(i),\text{MA}}, \quad &
  O_3 &= \mathcal{O}_{f7}^{(i),\text{MA}}, \quad & O_4 &= \mathcal{O}_{f8}^{(i),\text{MA}}, \nonumber
 \end{alignat}
 of the representation $\tau^{D\underline{12}}_2$ we find the anomalous dimension matrix
 \begin{align}
  \gamma^\Gamma = \frac{\alpha_s}{4\pi} \left(
  \begin{array}{cccc}
   2\,\xi -\frac{22}{9}\, & -\frac{16}{9} & \frac{8}{9} & 0 \\[2mm]
   -\frac{8}{9} & 2\,\xi -\frac{14}{9}\, & -\frac{2}{3} & 0 \\[2mm]
   0 & 0 & 2\,\xi -\frac{26}{9}\, & 0 \\[2mm]
   0 & 0 & 0 & 2\,\xi -4\,
  \end{array}
  \right).
 \end{align}

\subsection{Operators with Two Derivatives}
 Finally, let us quote the results for operators with two covariant derivatives.
 In the $\overline{\text{H}(4)}$-irreducible representation $\tau^{DD\underline{4}}_1$ the operators 
 \begin{alignat}{3}
  O_1 &= \mathcal{O}_{ff1}^{(i),\text{MA}}, \quad &  O_2 &= \mathcal{O}_{ff2}^{(i),\text{MA}}, \quad &
  O_3 &= \mathcal{O}_{ff3}^{(i),\text{MA}}, \nonumber\\  O_4 &= \mathcal{O}_{gh1}^{(i),\text{MA}}, \quad &
  O_5 &= \mathcal{O}_{gh2}^{(i),\text{MA}}, \quad &  O_6 &= \mathcal{O}_{gh3}^{(i),\text{MA}}, \nonumber
 \end{alignat}
 form a basis and have the anomalous dimension matrix
 \begin{align}
  \gamma^\Gamma = \frac{\alpha_s}{4\pi} \left(
  \begin{array}{cccccc}
   2\, \xi -\frac{32}{9}\, & \frac{2}{3} & -\frac{1}{3} & 0 & \frac{10}{3} & -\frac{5}{3} \\[2mm]
   \frac{1}{3} & 2\, \xi -3\, & -\frac{2}{9} & \frac{4}{3} & \frac{5}{3} & -\frac{5}{3} \\[2mm]
   0 & 0 & 2\, \xi -\frac{31}{9}\, & 0 & 0 & -\frac{5}{3} \\[2mm]
   0 & \frac{8}{9} & -\frac{4}{9} & 2\, \xi -\frac{34}{9}\, & \frac{16}{9} & -\frac{8}{9} \\[2mm]
   \frac{5}{9} & \frac{5}{9} & -\frac{5}{9} & \frac{8}{9} & 2\, \xi -\frac{29}{9}\, & -\frac{2}{3} \\[2mm]
   0 & 0 & -\frac{5}{9} & 0 & 0 & 2\, \xi -\frac{41}{9}\,
  \end{array}
  \right).
 \end{align}
 The same anomalous dimension matrix appears for the operators 
 \begin{alignat}{3}
  O_1 &= \mathcal{O}_{ff4}^{(i),\text{MA}}, \quad & O_2 &= \mathcal{O}_{ff5}^{(i),\text{MA}}, \quad &
  O_3 &= \mathcal{O}_{ff6}^{(i),\text{MA}}, \nonumber\\ 
  O_4 &= \mathcal{O}_{gh4}^{(i),\text{MA}}, \quad & O_5 &= \mathcal{O}_{gh5}^{(i),\text{MA}}, \quad & O_6 &= \mathcal{O}_{gh6}^{(i),\text{MA}}, \nonumber
 \end{alignat}
 belonging to $\tau^{DD\underline 4}_2$. The representation $\tau^{DD\underline 8}$ with the operators 
 \begin{alignat}{3}
  O_1 &= \mathcal{O}_{ff7}^{(i),\text{MA}}, \quad &  O_2 &= \mathcal{O}_{ff8}^{(i),\text{MA}}, \quad &
  O_3 &= \mathcal{O}_{ff9}^{(i),\text{MA}}, \nonumber\\ O_4 &= \mathcal{O}_{gh7}^{(i),\text{MA}}, \quad &
  O_5 &= \mathcal{O}_{gh8}^{(i),\text{MA}}, \quad &  O_6 &= \mathcal{O}_{gh9}^{(i),\text{MA}}, \nonumber
 \end{alignat} 
 has the following anomalous dimension:
 \begin{align}
  \gamma^\Gamma = \frac{\alpha_s}{4\pi} \left(
  \begin{array}{cccccc}
   2\, \xi -\frac{32}{9}\, & -\frac{2}{3} & \frac{1}{3} & 0 & -\frac{10}{3} & \frac{5}{3} \\[2mm]
   -\frac{1}{3} & 2\, \xi -3\, & -\frac{2}{9} & -\frac{4}{3} & \frac{5}{3} & -\frac{5}{3} \\[2mm]
   0 & 0 & 2\, \xi -\frac{31}{9}\, & 0 & 0 & -\frac{5}{3} \\[2mm]
   0 & -\frac{8}{9} & \frac{4}{9} & 2\, \xi -\frac{34}{9}\, & -\frac{16}{9} & \frac{8}{9} \\[2mm]
   -\frac{5}{9} & \frac{5}{9} & -\frac{5}{9} & -\frac{8}{9} & 2\, \xi -\frac{29}{9}\, & -\frac{2}{3} \\[2mm]
   0 & 0 & -\frac{5}{9} & 0 & 0 & 2\, \xi -\frac{41}{9}\,
  \end{array}
  \right).
 \end{align}
 The irreducible representation $\tau^{DD\underline{12}}_1$ and its operator basis 
 \begin{alignat}{4}
  O_1 &= \mathcal{O}_{ff10}^{(i),\text{MA}}, &\quad O_2 &= \mathcal{O}_{ff11}^{(i),\text{MA}}, &\quad
  O_3 &= \mathcal{O}_{ff12}^{(i),\text{MA}}, &\quad O_4 &= \mathcal{O}_{ff13}^{(i),\text{MA}}, \nonumber\\
  O_5 &= \mathcal{O}_{gh10}^{(i),\text{MA}}, &\quad O_6 &= \mathcal{O}_{gh11}^{(i),\text{MA}}, &\quad
  O_7 &= \mathcal{O}_{gh12}^{(i),\text{MA}}, &\quad O_8 &= \mathcal{O}_{gh13}^{(i),\text{MA}}, \nonumber
 \end{alignat}
 has the one-loop anomalous dimension
 \begin{align} 
  \gamma^\Gamma = \frac{\alpha_s}{4\pi} \left(
   \begin{array}{cccccccc}
   2 \xi -\frac{32}{9}\, & \frac{2}{3} & -\frac{1}{3} & 0 & 0 & \frac{10}{3} & -\frac{5}{3} & 0 \\[2mm]
   \frac{1}{3} & 2 \xi -3\, & -\frac{2}{9} & 0 & \frac{4}{3} & \frac{5}{3} & -\frac{5}{3} & 0 \\[2mm]
   0 & 0 & 2 \xi -\frac{31}{9}\, & 0 & 0 & 0 & -\frac{5}{3} & 0 \\[2mm]
   0 & 0 & 0 & 2 \xi -\frac{40}{9}\, & 0 & 0 & 0 & -\frac{4}{3} \\[2mm]
   0 & \frac{8}{9} & -\frac{4}{9} & 0 & 2 \xi -\frac{34}{9}\, & \frac{16}{9} & -\frac{8}{9} & 0 \\[2mm]
   \frac{5}{9} & \frac{5}{9} & -\frac{5}{9} & 0 & \frac{8}{9} & 2 \xi -\frac{29}{9}\, & -\frac{2}{3} & 0 \\[2mm]
   0 & 0 & -\frac{5}{9} & 0 & 0 & 0 & 2 \xi -\frac{41}{9}\, & 0 \\[2mm]
   0 & 0 & 0 & -\frac{4}{9} & 0 & 0 & 0 & 2 \xi -\frac{16}{3}\,
  \end{array}
  \right).
 \end{align}
 The same anomalous dimension can be shown to apply to the second 
twelve-di\-men\-sion\-al representation $\tau^{DD\underline{12}}_2$ and 
its operator basis 
 \begin{alignat}{4}
  O_1 &= \mathcal{O}_{ff14}^{(i),\text{MA}}, &\quad O_2 &= \mathcal{O}_{ff15}^{(i),\text{MA}}, &\quad O_3 &= \mathcal{O}_{ff16}^{(i),\text{MA}}, &\quad O_4 &= \mathcal{O}_{ff17}^{(i),\text{MA}}, \nonumber\\
  O_5 &= \mathcal{O}_{gh14}^{(i),\text{MA}}, &\quad O_6 &= \mathcal{O}_{gh15}^{(i),\text{MA}}, &\quad O_7 &= \mathcal{O}_{gh16}^{(i),\text{MA}}, &\quad O_8 &= \mathcal{O}_{gh17}^{(i),\text{MA}}. \nonumber
 \end{alignat}

\section{The Renormalization Matrices for Three-Quark Operators}
\label{App:MSRenormalization}
 In this appendix we present the renormalization matrix $Z(2\,\text{GeV})$ and the error matrices $E^{\text{sy}}$. As the statistical errors are much smaller than the systematic uncertainties we do not quote them here. 
Note that the error due to the error of $\Lambda^{\overline{\text{MS}}}$
leads to an additional non-negligible uncertainty (similar in size
to $E^{\text{sy}}$) for operators with two derivatives.
We also list the operator bases for all irreducible representations. For the notation of the three-quark operators compare again \cite{Kaltenbrunner:2008pb}. The renormalized operators are related to the bare lattice operators by
 \begin{align}
  O_i^{\overline{\text{MS}}} = Z_{ij} O_j.
 \end{align}
 
 We have results not only for the two lattices presented in the following, but also for all other lattices in Table \ref{tab:lattices}. However, in order to keep the paper at a reasonable length we restrict ourselves to the largest lattices ($24^3\times 48$) at $\beta=5.29$ and $\beta=5.40$.

\subsection{Operators without Derivatives in Representation $\tau^{\underline{4}}_1$}
This irreducible representation contains two mixing multiplets. The renormalization matrix $Z_{ij}$ is given in the following operator basis:
\begin{align}
O_1 = \mathcal{O}_1^{(i),\text{MA}}, &\quad O_2 = \mathcal{O}_3^{(i),\text{MA}}.
\end{align}
We now present our chirally extrapolated results.

$\beta=5.29$, lattice size: $24^3\times48$
\begin{align}
Z &= \left(
  \begin{array}{cc}
   0.6838 & -0.0290 \\
   -0.0066 & 0.6901
  \end{array}
  \right), \nonumber\\
E^{\text{sy}} &= \left(
  \begin{array}{cc}
   0.0176 & 0.0095 \\
   0.0021 & 0.0190
  \end{array}
  \right).
\end{align}

$\beta=5.40$, lattice size: $24^3\times48$
\begin{align}
Z &= \left(
  \begin{array}{cc}
   0.6892 & -0.0285 \\
   -0.0065 & 0.6953
  \end{array}
  \right), \nonumber\\
E^{\text{sy}} &= \left(
  \begin{array}{cc}
   0.0151 & 0.0083 \\
   0.0020 & 0.0163
  \end{array}
  \right).
\end{align}

\subsection{Operators without Derivatives in Representation $\tau^{\underline{12}}_1$}
Only one operator multiplet belongs to $\tau^{\underline{12}}_1$. Hence, there is no mixing present and the operator basis is given by
\begin{align}
O_1 &= \mathcal{O}_7^{(i),\text{MA}}.
\end{align}

$\beta=5.29$, lattice size: $24^3\times48$
\begin{align}
Z &= \left(
  \begin{array}{c}
   0.8047
  \end{array}
  \right), \nonumber\\
E^{\text{sy}} &= \left(
  \begin{array}{c}
   0.0176
  \end{array}
  \right).
\end{align}

$\beta=5.40$, lattice size: $24^3\times48$
\begin{align}
Z &= \left(
  \begin{array}{c}
   0.8131
  \end{array}
  \right), \nonumber\\
E^{\text{sy}} &= \left(
  \begin{array}{c}
   0.0139
  \end{array}
  \right).
\end{align}

\subsection{Operators with One Derivative in Representation $\tau^{\underline{8}}$}
In leading twist there is also no mixing for this representation. We take the basis
\begin{align}
O_1 &= \mathcal{O}_{f1}^{(i),\text{MA}}.
\end{align}

$\beta=5.29$, lattice size: $24^3\times48$
\begin{align}
Z &= \left(
  \begin{array}{c}
   1.1080
  \end{array}
  \right), \nonumber\\
E^{\text{sy}} &= \left(
  \begin{array}{c}
   0.0164
  \end{array}
  \right).
\end{align}

$\beta=5.40$, lattice size: $24^3\times48$
\begin{align}
Z &= \left(
  \begin{array}{c}
   1.1260
  \end{array}
  \right), \nonumber\\
E^{\text{sy}} &= \left(
  \begin{array}{c}
   0.0172
  \end{array}
  \right).
\end{align}

\subsection{Operators with One Derivative in Representation $\tau^{\underline{12}}_1$}
There are four mixing multiplets and we take
\begin{align}
  O_1 = \mathcal{O}_{f2}^{(i),\text{MA}}, \quad & O_2 = \mathcal{O}_{f3}^{(i),\text{MA}}, \nonumber\\
  O_3 = \mathcal{O}_{f4}^{(i),\text{MA}}, \quad & O_4 = \frac{1}{a} \, \mathcal{O}_7^{(i),\text{MA}}.
\end{align}

$\beta=5.29$, lattice size: $24^3\times48$
\begin{align}
Z &= \left(
  \begin{array}{cccc}
   1.0440 & 0.0989 & -0.0653 & 9.1\times 10^{\text{-4}} \\
   0.0525 & 0.9841 & 0.0452 & -6.6\times 10^{\text{-5}} \\
   0.0033 & -0.0018 & 1.0730 & -2.5\times 10^{\text{-4}}
  \end{array}
  \right), \nonumber\\
E^{\text{sy}} &= \left(
  \begin{array}{cccc}
   0.0085 & 0.0190 & 0.0106 & 1.8\times 10^{\text{-4}} \\
   0.0064 & 0.0127 & 0.0038 & 1.7\times 10^{\text{-4}} \\
   0.0013 & 0.0014 & 0.0136 & 1.6\times 10^{\text{-4}}
  \end{array}
  \right).
\end{align}

$\beta=5.40$, lattice size: $24^3\times48$
\begin{align}
Z &= \left(
  \begin{array}{cccc}
   1.0540 & 0.1081 & -0.0693 & 9.3\times 10^{\text{-4}} \\
   0.0564 & 0.9920 & 0.0483 & -2.0\times 10^{\text{-5}} \\
   0.0033 & -0.0028 & 1.0890 & -2.1\times 10^{\text{-4}}
  \end{array}
  \right), \nonumber\\
E^{\text{sy}} &= \left(
  \begin{array}{cccc}
   0.0062 & 0.0185 & 0.0095 & 1.5\times 10^{\text{-4}} \\
   0.0061 & 0.0087 & 0.0045 & 1.5\times 10^{\text{-4}} \\
   0.0010 & 0.0017 & 0.0150 & 1.4\times 10^{\text{-4}}
  \end{array}
  \right).
\end{align}

\subsection{Operators with One Derivative in Representation $\tau^{\underline{12}}_2$}
We work in the following operator basis:
\begin{align}
  O_1 = \mathcal{O}_{f5}^{(i),\text{MA}}, \quad & O_2 = \mathcal{O}_{f6}^{(i),\text{MA}}, \nonumber\\
  O_3 = \mathcal{O}_{f7}^{(i),\text{MA}}, \quad & O_4 = \mathcal{O}_{f8}^{(i),\text{MA}}. 
\end{align}

$\beta=5.29$, lattice size: $24^3\times48$
\begin{align}
Z &= \left(
  \begin{array}{cccc}
   1.0350 & 0.0975 & -0.0632 & -0.0011 \\
   0.0502 & 0.9813 & 0.0456 & -6.7\times 10^{\text{-4}} \\
   0.0080 & -0.0065 & 1.0720 & 9.1\times 10^{\text{-4}} \\
   -4.8\times 10^{\text{-4}} & -0.0022 & 0.0109 & 1.1150
  \end{array}
  \right), \nonumber\\
E^{\text{sy}} &= \left(
  \begin{array}{cccc}
   0.0091 & 0.0171 & 0.0107 & 0.0014 \\
   0.0082 & 0.0135 & 0.0050 & 7.0\times 10^{\text{-4}} \\
   0.0016 & 8.0\times 10^{\text{-4}} & 0.0138 & 4.9\times 10^{\text{-4}} \\
   8.3\times 10^{\text{-4}} & 0.0015 & 6.9\times 10^{\text{-4}} & 0.0174
  \end{array}
  \right).
\end{align}

$\beta=5.40$, lattice size: $24^3\times48$
\begin{align}
Z &= \left(
  \begin{array}{cccc}
   1.0470 & 0.1066 & -0.0675 & -0.0013 \\
   0.0544 & 0.9898 & 0.0487 & -7.8\times 10^{\text{-4}} \\
   0.0080 & -0.0064 & 1.0870 & 8.4\times 10^{\text{-4}} \\
   -6.4\times 10^{\text{-4}} & -0.0026 & 0.0111 & 1.1320
  \end{array}
  \right), \nonumber\\
E^{\text{sy}} &= \left(
  \begin{array}{cccc}
   0.0093 & 0.0166 & 0.0096 & 0.0010 \\
   0.0076 & 0.0094 & 0.0054 & 6.1\times 10^{\text{-4}} \\
   9.8\times 10^{\text{-4}} & 6.4\times 10^{\text{-4}} & 0.0146 & 3.8\times \
  10^{\text{-4}} \\
   5.8\times 10^{\text{-4}} & 0.0013 & 8.0\times 10^{\text{-4}} & 0.0176
  \end{array}
  \right).
\end{align}

\subsection{Operators with Two Derivatives in Representation $\tau^{\underline{4}}_1$}
Here, mixing with lower-dimensional operators occurs:
\begin{alignat}{4}
  O_1 &= \mathcal{O}_{ff1}^{(i),\text{MA}}, \quad &  O_2 &= \mathcal{O}_{ff2}^{(i),\text{MA}}, \quad &
  O_3 &= \mathcal{O}_{ff3}^{(i),\text{MA}}, \quad &  O_4 &= \mathcal{O}_{gh1}^{(i),\text{MA}}, \nonumber\\
  O_5 &= \mathcal{O}_{gh2}^{(i),\text{MA}}, \quad &  O_6 &= \mathcal{O}_{gh3}^{(i),\text{MA}}, \quad &
  O_7 &= \frac{1}{a^2} \cdot \mathcal{O}_1^{(i),\text{MA}}, \quad &  O_8 &= \frac{1}{a^2} \, \mathcal{O}_3^{(i),\text{MA}}.
\end{alignat}

$\beta=5.29$, lattice size: $24^3\times48$
\begin{scriptsize}
\begin{align}
Z &= \left(
  \begin{array}{cccccccc}
   1.3280 & 0.0296 & -0.0063 & 0.0281 & -0.1538 & 0.0644 & 6.1\times \
  10^{\text{-4}} & 7.3\times 10^{\text{-4}} \\
   0.0142 & 1.2900 & -0.0070 & -0.0772 & -0.0416 & 0.0712 & -3.3\times \
  10^{\text{-4}} & -0.0034 \\
   2.2\times 10^{\text{-4}} & 0.0055 & 1.2640 & -0.0044 & -0.0014 & 0.1108 & \
  5.9\times 10^{\text{-5}} & -0.0080 \\
   0.0228 & -0.0754 & 0.0427 & 1.2930 & -0.0557 & 0.0339 & 0.0011 & 7.2\times \
  10^{\text{-4}} \\
   -0.0809 & -0.0729 & 0.0667 & -0.0473 & 1.2040 & 0.0536 & 4.0\times \
  10^{\text{-4}} & -0.0014 \\
   0.0296 & 0.0536 & 0.0499 & 0.0330 & 0.0721 & 1.2560 & 3.1\times \
  10^{\text{-5}} & -0.0053
  \end{array}
  \right), \nonumber\\
E^{\text{sy}} &= \left(
  \begin{array}{cccccccc}
   0.0341 & 0.0165 & 0.0125 & 0.0163 & 0.0269 & 0.0101 & 3.2\times \
  10^{\text{-4}} & 3.7\times 10^{\text{-4}} \\
   0.0210 & 0.0227 & 0.0075 & 0.0123 & 0.0265 & 0.0173 & 3.1\times \
  10^{\text{-4}} & 0.0025 \\
   0.0072 & 0.0041 & 0.0148 & 0.0111 & 0.0055 & 0.0041 & 8.3\times \
  10^{\text{-5}} & 0.0057 \\
   0.0203 & 0.0479 & 0.0123 & 0.0026 & 0.0733 & 0.0395 & 5.5\times \
  10^{\text{-4}} & 2.9\times 10^{\text{-4}} \\
   0.0278 & 0.0273 & 0.0162 & 0.0238 & 0.0193 & 0.0249 & 4.7\times \
  10^{\text{-5}} & 0.0033 \\
   0.0137 & 0.0137 & 0.0134 & 0.0123 & 0.0310 & 0.0469 & 1.3\times \
  10^{\text{-4}} & 0.0052
  \end{array}
  \right).
\end{align}
\end{scriptsize}

$\beta=5.40$, lattice size: $24^3\times48$
\begin{scriptsize}
\begin{align}
Z &= \left(
  \begin{array}{cccccccc}
   1.3390 & 0.0282 & -0.0010 & 0.0306 & -0.1620 & 0.0693 & 6.6\times \
  10^{\text{-4}} & 6.4\times 10^{\text{-4}} \\
   0.0167 & 1.2950 & -0.0030 & -0.0808 & -0.0458 & 0.0750 & -2.4\times \
  10^{\text{-4}} & -0.0040 \\
   0.0022 & 0.0058 & 1.2710 & -0.0019 & -1.4\times 10^{\text{-4}} & 0.1167 & \
  7.4\times 10^{\text{-5}} & -0.0092 \\
   0.0174 & -0.0892 & 0.0468 & 1.3010 & -0.0803 & 0.0464 & 8.7\times \
  10^{\text{-4}} & 8.3\times 10^{\text{-4}} \\
   -0.0872 & -0.0794 & 0.0708 & -0.0564 & 1.2080 & 0.0615 & 4.0\times \
  10^{\text{-4}} & -0.0020 \\
   0.0249 & 0.0475 & 0.0550 & 0.0285 & 0.0618 & 1.2810 & 2.8\times \
  10^{\text{-5}} & -0.0063
  \end{array}
  \right), \nonumber\\
E^{\text{sy}} &= \left(
  \begin{array}{cccccccc}
   0.0372 & 0.0205 & 0.0099 & 0.0196 & 0.0327 & 0.0134 & 3.6\times \
  10^{\text{-4}} & 4.3\times 10^{\text{-4}} \\
   0.0228 & 0.0179 & 0.0070 & 0.0123 & 0.0303 & 0.0195 & 3.6\times \
  10^{\text{-4}} & 0.0023 \\
   0.0071 & 0.0052 & 0.0098 & 0.0108 & 0.0058 & 0.0031 & 7.3\times \
  10^{\text{-5}} & 0.0055 \\
   0.0208 & 0.0493 & 0.0138 & 0.0033 & 0.0732 & 0.0390 & 5.7\times \
  10^{\text{-4}} & 3.6\times 10^{\text{-4}} \\
   0.0273 & 0.0281 & 0.0162 & 0.0246 & 0.0196 & 0.0243 & 5.9\times \
  10^{\text{-5}} & 0.0032 \\
   0.0138 & 0.0147 & 0.0139 & 0.0120 & 0.0304 & 0.0452 & 7.7\times \
  10^{\text{-5}} & 0.0051
  \end{array}
  \right).
\end{align}
\end{scriptsize}

\subsection{Operators with Two Derivatives in Representation $\tau^{\underline{4}}_2$}
This is the only irreducible representation of leading-twist operators with two derivatives that is not subject to mixing with lower-dimensional operators on the lattice:
\begin{alignat}{3}
O_1 &= \mathcal{O}_{ff4}^{(i),\text{MA}}, \quad & O_2 &= \mathcal{O}_{ff5}^{(i),\text{MA}}, \quad &
O_3 &= \mathcal{O}_{ff6}^{(i),\text{MA}}, \nonumber\\ 
O_4 &= \mathcal{O}_{gh4}^{(i),\text{MA}}, \quad & O_5 &= \mathcal{O}_{gh5}^{(i),\text{MA}}, \quad & O_6 &= \mathcal{O}_{gh6}^{(i),\text{MA}}.
\end{alignat}

$\beta=5.29$, lattice size: $24^3\times48$
\begin{scriptsize}
\begin{align}
Z &= \left(
  \begin{array}{cccccc}
   1.2810 & -0.0332 & 0.0047 & 0.0284 & -0.2071 & 0.1238 \\
   -0.0204 & 1.2450 & 0.0051 & -0.0922 & -0.1045 & 0.1201 \\
   -0.0080 & -0.0125 & 1.2710 & -8.8\times 10^{\text{-4}} & -0.0123 & 0.1309 \
  \\
   0.0367 & -0.0044 & 0.0288 & 1.3540 & -0.0408 & -0.0173 \\
   -0.0157 & -0.0213 & 0.0235 & -0.0610 & 1.3260 & 0.0319 \\
   0.0322 & 0.0242 & 0.0269 & 0.0049 & 0.0622 & 1.3480
  \end{array}
  \right), \nonumber\\
E^{\text{sy}} &= \left(
  \begin{array}{cccccc}
   0.0022 & 0.0467 & 0.0229 & 0.0165 & 0.0363 & 0.0418 \\
   0.0260 & 0.0379 & 0.0222 & 0.0071 & 0.0442 & 0.0476 \\
   0.0050 & 0.0053 & 0.0142 & 0.0093 & 0.0195 & 0.0057 \\
   0.0098 & 0.0280 & 0.0077 & 0.0243 & 0.0566 & 0.0362 \\
   0.0079 & 0.0031 & 0.0041 & 0.0150 & 0.0170 & 0.0187 \\
   0.0099 & 0.0100 & 0.0063 & 0.0042 & 0.0205 & 0.0683
  \end{array}
  \right).
\end{align}
\end{scriptsize}

$\beta=5.40$, lattice size: $24^3\times48$
\begin{scriptsize}
\begin{align}
Z &= \left(
  \begin{array}{cccccc}
   1.2830 & -0.0503 & 0.0142 & 0.0296 & -0.2268 & 0.1382 \\
   -0.0287 & 1.2360 & 0.0137 & -0.0994 & -0.1214 & 0.1365 \\
   -0.0065 & -0.0132 & 1.2820 & 3.8\times 10^{\text{-4}} & -0.0089 & 0.1363 \\
   0.0329 & -0.0138 & 0.0310 & 1.3690 & -0.0637 & -0.0021 \\
   -0.0193 & -0.0231 & 0.0253 & -0.0679 & 1.3330 & 0.0412 \\
   0.0291 & 0.0207 & 0.0301 & 0.0035 & 0.0547 & 1.3770
  \end{array}
  \right), \nonumber\\
E^{\text{sy}} &= \left(
  \begin{array}{cccccc}
   0.0053 & 0.0525 & 0.0249 & 0.0118 & 0.0464 & 0.0477 \\
   0.0281 & 0.0416 & 0.0245 & 0.0101 & 0.0497 & 0.0508 \\
   0.0040 & 0.0056 & 0.0153 & 0.0066 & 0.0145 & 0.0081 \\
   0.0099 & 0.0257 & 0.0061 & 0.0283 & 0.0539 & 0.0366 \\
   0.0070 & 0.0020 & 0.0029 & 0.0134 & 0.0176 & 0.0186 \\
   0.0093 & 0.0103 & 0.0063 & 0.0042 & 0.0209 & 0.0682
  \end{array}
  \right).
\end{align}
\end{scriptsize}

\subsection{Operators with Two Derivatives in Representation $\tau^{\underline{8}}$}
This irreducible representation mixes with one lower-dimensional operator. The basis is
\begin{alignat}{4}
 O_1 &= \mathcal{O}_{ff7}^{(i),\text{MA}}, \quad &  O_2 &= \mathcal{O}_{ff8}^{(i),\text{MA}}, \quad &
 O_3 &= \mathcal{O}_{ff9}^{(i),\text{MA}}, \quad &  O_4 &= \mathcal{O}_{gh7}^{(i),\text{MA}}, \nonumber\\
 O_5 &= \mathcal{O}_{gh8}^{(i),\text{MA}}, \quad &  O_6 &= \mathcal{O}_{gh9}^{(i),\text{MA}}, \quad &
 O_7 &= \frac{1}{a} \, \mathcal{O}_{f1}^{(i),\text{MA}}. \quad & & 
\end{alignat}

$\beta=5.29$, lattice size: $24^3\times48$
\begin{scriptsize}
\begin{align}
  Z &= \left(
  \begin{array}{ccccccc}
   1.3080 & 0.0156 & -0.0055 & 0.0165 & 0.1784 & -0.0734 & 0.0054 \\
   -0.0113 & 1.2590 & 5.8\times 10^{\text{-4}} & 0.0785 & -0.0553 & 0.0638 & \
  -0.0020 \\
   -0.0069 & -0.0177 & 1.2620 & 0.0033 & -0.0130 & 0.1216 & -4.6\times \
  10^{\text{-4}} \\
   0.0426 & 0.0415 & -0.0296 & 1.3140 & 0.0541 & -0.0238 & -0.0056 \\
   0.0635 & -0.0557 & 0.0551 & 0.0487 & 1.2000 & 0.0841 & -5.4\times \
  10^{\text{-4}} \\
   -0.0471 & 0.0606 & 0.0468 & -0.0322 & 0.0848 & 1.2620 & -8.7\times \
  10^{\text{-4}}
  \end{array}
  \right), \nonumber\\
  E^{\text{sy}} &= \left(
  \begin{array}{ccccccc}
   0.0190 & 0.0300 & 0.0345 & 0.0119 & 0.0095 & 0.0099 & 0.0039 \\
   0.0085 & 0.0300 & 0.0328 & 0.0204 & 0.0150 & 0.0106 & 4.3\times \
  10^{\text{-4}} \\
   0.0060 & 6.1\times 10^{\text{-4}} & 0.0135 & 0.0097 & 0.0083 & 0.0058 & \
  5.1\times 10^{\text{-4}} \\
   0.0163 & 0.0205 & 0.0028 & 0.0259 & 0.0507 & 0.0422 & 0.0011 \\
   0.0148 & 0.0076 & 0.0032 & 0.0047 & 0.0219 & 0.0285 & 0.0013 \\
   0.0101 & 0.0086 & 0.0113 & 0.0050 & 0.0248 & 0.0489 & 5.2\times \
  10^{\text{-4}}
  \end{array}
  \right).
\end{align}
\end{scriptsize}

$\beta=5.40$, lattice size: $24^3\times48$ 
\begin{scriptsize}
\begin{align}
  Z &= \left(
    \begin{array}{ccccccc}
     1.3150 & 0.0220 & -0.0133 & 0.0139 & 0.1879 & -0.0742 & 0.0067 \\
     -0.0124 & 1.2560 & 0.0090 & 0.0882 & -0.0621 & 0.0653 & -0.0020 \\
     -0.0084 & -0.0176 & 1.2680 & 0.0017 & -0.0123 & 0.1278 & -6.6\times \
    10^{\text{-4}} \\
     0.0349 & 0.0517 & -0.0295 & 1.3240 & 0.0809 & -0.0426 & -0.0053 \\
     0.0691 & -0.0584 & 0.0565 & 0.0543 & 1.2020 & 0.0944 & -1.6\times \
    10^{\text{-4}} \\
     -0.0410 & 0.0549 & 0.0507 & -0.0291 & 0.0715 & 1.2890 & -8.2\times \
    10^{\text{-4}}
    \end{array}
    \right), \nonumber\\
  E^{\text{sy}} &= \left(
    \begin{array}{ccccccc}
     0.0207 & 0.0371 & 0.0389 & 0.0154 & 0.0148 & 0.0080 & 0.0037 \\
     0.0073 & 0.0303 & 0.0387 & 0.0213 & 0.0194 & 0.0095 & 5.2\times \
    10^{\text{-4}} \\
     0.0050 & 0.0026 & 0.0171 & 0.0076 & 0.0040 & 0.0077 & 3.8\times \
    10^{\text{-4}} \\
     0.0141 & 0.0172 & 0.0042 & 0.0291 & 0.0447 & 0.0371 & 0.0016 \\
     0.0132 & 0.0053 & 0.0027 & 0.0040 & 0.0150 & 0.0250 & 0.0015 \\
     0.0095 & 0.0087 & 0.0118 & 0.0053 & 0.0245 & 0.0485 & 6.4\times \
    10^{\text{-4}}
    \end{array}
    \right).
\end{align}
\end{scriptsize}

\subsection{Operators with Two Derivatives in Representation $\tau^{\underline{12}}_1$}
Here, twelve multiplets of operators mix with each other under renormalization. Four of them have lower dimension:
\begin{alignat}{4}
O_1 &= \mathcal{O}_{ff10}^{(i),\text{MA}}, &\quad O_2 &= \mathcal{O}_{ff11}^{(i),\text{MA}}, &\quad
O_3 &= \mathcal{O}_{ff12}^{(i),\text{MA}}, &\quad O_4 &= \mathcal{O}_{ff13}^{(i),\text{MA}}, \nonumber\\
O_5 &= \mathcal{O}_{gh10}^{(i),\text{MA}}, &\quad O_6 &= \mathcal{O}_{gh11}^{(i),\text{MA}}, &\quad
O_7 &= \mathcal{O}_{gh12}^{(i),\text{MA}}, &\quad O_8 &= \mathcal{O}_{gh13}^{(i),\text{MA}}, \nonumber\\
O_9 &= \frac{1}{a} \, \mathcal{O}_{f2}^{(i),\text{MA}},   &\quad O_{10} &= \frac{1}{a} \, \mathcal{O}_{f3}^{(i),\text{MA}}, &\quad
O_{11} &= \frac{1}{a} \, \mathcal{O}_{f4}^{(i),\text{MA}},  &\quad O_{12} &= \frac{1}{a^2} \, \mathcal{O}_7^{(i),\text{MA}}.  \nonumber
\end{alignat}
 We have split off the last four columns of the renormalization matrix, which describe the mixing with the lower-dimensional operators $O_9$,...,$O_{12}$, and display the related coefficients in a separate matrix $Z'$.

$\beta=5.29$, lattice size: $24^3\times48$
\begin{scriptsize}
\begin{align}
Z &= \left(
  \begin{array}{cccccccccccc}
   1.3020 & -0.0310 & 0.0186 & 0.0039 & 0.0223 & -0.1658 & 0.0771 & -0.0010  \\
   -0.0162 & 1.2540 & 0.0113 & -4.3\times 10^{\text{-4}} & -0.0941 & -0.0925 & \
  0.1026 & 8.0\times 10^{\text{-4}}  \\
   0.0130 & 0.0171 & 1.2710 & -6.1\times 10^{\text{-4}} & 0.0206 & 0.0630 & \
  0.0720 & -3.3\times 10^{\text{-4}}  \\
   8.6\times 10^{\text{-4}} & -0.0057 & 0.0024 & 1.3150 & -0.0095 & -0.0177 & \
  0.0080 & 0.1001  \\
   0.0771 & 0.0010 & 0.0308 & 8.9\times 10^{\text{-4}} & 1.3350 & 0.0166 & \
  -0.0444 & -3.4\times 10^{\text{-4}} \\
   -0.0306 & -0.0114 & 0.0310 & -1.7\times 10^{\text{-4}} & -0.0403 & 1.2920 & \
  0.0306 & -9.1\times 10^{\text{-4}}  \\
   0.0646 & 0.0720 & 0.0436 & 6.6\times 10^{\text{-4}} & 0.0227 & 0.1103 & \
  1.2640 & 2.0\times 10^{\text{-4}}  \\
   -0.0010 & 0.0085 & -0.0043 & 0.0120 & 0.0064 & 0.0027 & -0.0021 & 1.3330 
  \end{array}
  \right), \nonumber\\
E^{\text{sy}} &= \left(
  \begin{array}{cccccccccccc}
   0.0216 & 0.0576 & 0.0399 & 0.0016 & 0.0116 & 0.0515 & 0.0352 & 5.1\times \
  10^{\text{-4}}  \\
   0.0138 & 0.0479 & 0.0329 & 2.2\times 10^{\text{-4}} & 0.0291 & 0.0459 & \
  0.0321 & 4.4\times 10^{\text{-4}}  \\
   0.0077 & 0.0191 & 0.0182 & 5.4\times 10^{\text{-4}} & 0.0041 & 0.0195 & \
  0.0249 & 3.2\times 10^{\text{-4}}  \\
   0.0015 & 0.0028 & 0.0014 & 0.0256 & 0.0018 & 0.0029 & 0.0026 & 0.0090  \\
   0.0029 & 0.0086 & 0.0067 & 3.5\times 10^{\text{-4}} & 0.0253 & 0.0378 & \
  0.0158 & 2.9\times 10^{\text{-4}}  \\
   0.0066 & 0.0030 & 0.0029 & 5.4\times 10^{\text{-4}} & 0.0152 & 0.0128 & \
  0.0177 & 4.9\times 10^{\text{-4}}  \\
   0.0045 & 0.0099 & 0.0104 & 9.8\times 10^{\text{-4}} & 0.0146 & 0.0203 & \
  0.0511 & 4.4\times 10^{\text{-4}}  \\
   8.7\times 10^{\text{-4}} & 0.0023 & 4.7\times 10^{\text{-4}} & 0.0042 & \
  0.0019 & 0.0013 & 0.0021 & 0.0325 
  \end{array}
  \right).
\end{align}
\end{scriptsize}
\begin{scriptsize}
\begin{align}
Z' &= \left(
  \begin{array}{cccccccccccc}
   -7.4\times 10^{\text{-4}} & -5.6\times 10^{\text{-4}} & 0.0051 & 0.0189 \\
   -0.0057 & 0.0048 & -0.0052 & 0.0128 \\
   -0.0019 & 0.0043 & -0.0028 & 0.0023 \\
    0.0070 & -0.0186 & 0.0088 & -3.1\times 10^{\text{-4}} \\
    0.0017 & -0.0033 & 0.0068 & -0.0249 \\
    -0.0049 & 0.0075 & -0.0089 & -0.0205 \\
    -0.0043 & 0.0024 & -0.0075 & -0.0088 \\
    0.0100 & -0.0041 & 6.2\times 10^{\text{-4}} & -0.0011
  \end{array}
  \right), \nonumber\\
E'^{\text{sy}} &= \left(
  \begin{array}{cccccccccccc}
   9.9\times 10^{\text{-4}} & 6.5\times 10^{\text{-4}} & \
  9.8\times 10^{\text{-4}} & 0.0142 \\
   0.0017 & 0.0055 & 0.0015 & 0.0097 \\
   5.4\times 10^{\text{-4}} & 0.0041 & 0.0021 & 0.0024 \\
   0.0048 & 0.0077 & 0.0049 & 2.6\times 10^{\text{-4}} \\
   0.0012 & 0.0029 & 0.0018 & 0.0061 \\
   0.0014 & 0.0027 & 0.0012 & 0.0071 \\
   0.0024 & 0.0034 & 0.0018 & 0.0016 \\
   0.0049 & 0.0052 & 0.0038 & 4.3\times 10^{\text{-4}}
  \end{array}
  \right).
\end{align}
\end{scriptsize}

$\beta=5.40$, lattice size: $24^3\times48$
\begin{scriptsize}
\begin{align}
Z &= \left(
  \begin{array}{cccccccccccc}
   1.3020 & -0.0515 & 0.0317 & 0.0039 & 0.0184 & -0.1943 & 0.0940 & -8.9\times \
  10^{\text{-4}} \\
   -0.0212 & 1.2450 & 0.0210 & -4.3\times 10^{\text{-4}} & -0.1064 & -0.1106 & \
  0.1174 & 7.9\times 10^{\text{-4}}  \\
   0.0112 & 0.0113 & 1.2820 & -5.7\times 10^{\text{-4}} & 0.0186 & 0.0579 & \
  0.0845 & -4.4\times 10^{\text{-4}} \\
   0.0013 & -0.0046 & 0.0019 & 1.3220 & -0.0093 & -0.0166 & 0.0071 & 0.1027 \\
   0.0749 & -0.0050 & 0.0329 & 7.6\times 10^{\text{-4}} & 1.3470 & -0.0019 & \
  -0.0359 & -3.2\times 10^{\text{-4}} \\
   -0.0335 & -0.0137 & 0.0328 & 1.0\times 10^{\text{-4}} & -0.0483 & 1.3000 & \
  0.0382 & -9.3\times 10^{\text{-4}}  \\
   0.0617 & 0.0676 & 0.0472 & 0.0010 & 0.0182 & 0.1031 & 1.2900 & 1.7\times \
  10^{\text{-4}}  \\
   -7.9\times 10^{\text{-4}} & 0.0089 & -0.0043 & 0.0158 & 0.0065 & 0.0027 & \
  -0.0024 & 1.3560 
  \end{array}
  \right), \nonumber\\
E^{\text{sy}} &= \left(
  \begin{array}{cccccccccccc}
   0.0192 & 0.0595 & 0.0395 & 9.2\times 10^{\text{-4}} & 0.0111 & 0.0533 & \
  0.0372 & 5.4\times 10^{\text{-4}} \\
   0.0132 & 0.0463 & 0.0328 & 5.4\times 10^{\text{-5}} & 0.0274 & 0.0433 & \
  0.0315 & 1.7\times 10^{\text{-4}}  \\
   0.0078 & 0.0189 & 0.0197 & 2.2\times 10^{\text{-4}} & 0.0051 & 0.0195 & \
  0.0253 & 4.0\times 10^{\text{-4}}  \\
   0.0014 & 0.0030 & 0.0011 & 0.0167 & 0.0011 & 0.0031 & 0.0030 & 0.0077  \\
   0.0015 & 0.0065 & 0.0042 & 4.0\times 10^{\text{-4}} & 0.0253 & 0.0322 & \
  0.0126 & 1.8\times 10^{\text{-4}}  \\
   0.0061 & 0.0014 & 0.0031 & 7.2\times 10^{\text{-4}} & 0.0146 & 0.0146 & \
  0.0159 & 3.4\times 10^{\text{-4}}  \\
   0.0044 & 0.0081 & 0.0096 & 0.0011 & 0.0130 & 0.0170 & 0.0481 & 3.4\times \
  10^{\text{-4}}  \\
   4.9\times 10^{\text{-4}} & 0.0021 & 5.5\times 10^{\text{-4}} & 0.0061 & \
  0.0015 & 8.7\times 10^{\text{-4}} & 0.0014 & 0.0310 
  \end{array}
  \right).
\end{align}
\begin{align}
Z' &= \left(
  \begin{array}{cccccccccccc}
   -9.4\times 10^{\text{-4}} & -4.5\times 10^{\text{-4}} & 0.0046 & 0.0208 \\
   -0.0059 & 0.0060 & -0.0051 & 0.0139 \\
   -0.0017 & 0.0052 & -0.0033 & 0.0025 \\
   0.0057 & -0.0188 & 0.0091 & -3.0\times 10^{\text{-4}} \\
   0.0020 & -0.0041 & 0.0070 & -0.0236 \\
   -0.0044 & 0.0066 & -0.0084 & -0.0200 \\
   -0.0046 & 0.0029 & -0.0075 & -0.0083 \\
   0.0101 & -0.0046 & -1.8\times 10^{\text{-4}} & -9.4\times 10^{\text{-4}}
  \end{array}
  \right), \nonumber\\
E'^{\text{sy}} &= \left(
  \begin{array}{cccccccccccc}
   0.0011 & 6.4\times 10^{\text{-4}} & 9.3\times 10^{\text{-4}} & 0.0106 \\
   0.0013 & 0.0047 & 8.4\times 10^{\text{-4}} & 0.0072 \\
   6.3\times 10^{\text{-4}} & 0.0033 & 0.0016 & 0.0014 \\
   0.0049 & 0.0053 & 0.0032 & 1.4\times 10^{\text{-4}} \\
   9.8\times 10^{\text{-4}} & 0.0027 & 0.0014 & 0.0061 \\
   0.0016 & 0.0029 & 0.0013 & 0.0038 \\
   0.0018 & 0.0025 & 0.0012 & 0.0020 \\
   0.0030 & 0.0034 & 0.0041 & 4.8\times 10^{\text{-4}}
  \end{array}
  \right).
\end{align}
\end{scriptsize}

\subsection{Operators with Two Derivatives in Representation $\tau^{\underline{12}}_2$}
Finally, we have the representation $\tau^{DD\underline{12}}_2$. As basis we take the operators
\begin{alignat}{4}
 O_1 &= \mathcal{O}_{ff14}^{(i),\text{MA}}, &\quad O_2 &= \mathcal{O}_{ff15}^{(i),\text{MA}}, &\quad O_3 &= \mathcal{O}_{ff16}^{(i),\text{MA}}, &\quad O_4 &= \mathcal{O}_{ff17}^{(i),\text{MA}}, \nonumber\\
 O_5 &= \mathcal{O}_{gh14}^{(i),\text{MA}}, &\quad O_6 &= \mathcal{O}_{gh15}^{(i),\text{MA}}, &\quad O_7 &= \mathcal{O}_{gh16}^{(i),\text{MA}}, &\quad O_8 &= \mathcal{O}_{gh17}^{(i),\text{MA}}, \nonumber\\
 O_9 &= \frac{1}{a} \cdot \mathcal{O}_{f5}^{(i),\text{MA}}, &\quad O_{10} &= \frac{1}{a} \cdot \mathcal{O}_{f6}^{(i),\text{MA}}, &\quad
 O_{11} &= \frac{1}{a} \cdot \mathcal{O}_{f7}^{(i),\text{MA}}, &\quad O_{12} &= \frac{1}{a} \cdot \mathcal{O}_{f8}^{(i),\text{MA}}. \nonumber
\end{alignat}
Again we split off the last columns of our renormalization matrix, which describe the mixing with the lower-dimensional operators $O_9$,...,$O_{12}$, and display the related coefficients in a separate matrix $Z'$.

$\beta=5.29$, lattice size: $24^3\times48$
\begin{scriptsize}
\begin{align}
Z &= \left(
  \begin{array}{cccccccccccc}
   1.3240 & -0.0052 & 0.0194 & -0.0020 & -0.0097 & -0.2074 & 0.0980 & 0.0063 \\
   -0.0060 & 1.2740 & -0.0029 & -4.8\times 10^{\text{-5}} & -0.0886 & -0.0937 \
  & 0.0962 & -1.6\times 10^{\text{-4}}  \\
   0.0113 & 0.0057 & 1.2760 & -7.2\times 10^{\text{-5}} & -0.0060 & 6.5\times \
  10^{\text{-5}} & 0.1001 & 3.6\times 10^{\text{-5}} \\
   0.0060 & 0.0097 & -0.0027 & 1.3210 & 0.0018 & 0.0073 & -0.0138 & 0.1163  \\
   0.0354 & -0.0538 & 0.0379 & 5.1\times 10^{\text{-5}} & 1.3290 & -0.0631 & \
  0.0136 & 0.0013 \\
   -0.0526 & -0.0480 & 0.0490 & 3.0\times 10^{\text{-4}} & -0.0532 & 1.2840 & \
  0.0509 & -0.0014  \\
   0.0314 & 0.0310 & 0.0553 & -2.0\times 10^{\text{-4}} & -2.3\times \
  10^{\text{-4}} & 0.0656 & 1.3130 & -2.3\times 10^{\text{-4}}  \\
   0.0051 & 2.2\times 10^{\text{-5}} & -8.2\times 10^{\text{-5}} & 0.0357 & \
  -0.0015 & 0.0045 & -0.0017 & 1.3470 
  \end{array}
  \right), \nonumber\\
E^{\text{sy}} &= \left(
  \begin{array}{cccccccccccc}
   0.0308 & 0.0110 & 0.0057 & 0.0012 & 0.0112 & 0.0406 & 0.0350 & 5.9\times \
  10^{\text{-4}}  \\
   0.0117 & 0.0147 & 0.0026 & 1.4\times 10^{\text{-4}} & 0.0055 & 0.0530 & \
  0.0447 & 4.2\times 10^{\text{-4}} \\
   0.0046 & 0.0050 & 0.0134 & 2.5\times 10^{\text{-4}} & 0.0035 & 0.0091 & \
  0.0144 & 2.7\times 10^{\text{-4}}  \\
   0.0013 & 0.0025 & 0.0014 & 0.0201 & 0.0016 & 0.0029 & 0.0042 & 0.0131  \\
   0.0154 & 0.0369 & 0.0135 & 2.0\times 10^{\text{-4}} & 0.0263 & 0.0694 & \
  0.0463 & 4.5\times 10^{\text{-4}}  \\
   0.0211 & 0.0272 & 0.0138 & 3.7\times 10^{\text{-4}} & 0.0170 & 0.0265 & \
  0.0450 & 6.7\times 10^{\text{-4}}  \\
   0.0065 & 0.0126 & 0.0099 & 3.0\times 10^{\text{-4}} & 0.0088 & 0.0180 & \
  0.0516 & 6.7\times 10^{\text{-4}}  \\
   0.0015 & 0.0015 & 6.8\times 10^{\text{-4}} & 0.0132 & 2.5\times \
  10^{\text{-4}} & 8.5\times 10^{\text{-4}} & 4.5\times 10^{\text{-4}} & \
  0.0423 
  \end{array}
  \right).
\end{align}
\begin{align}
Z' &= \left(
  \begin{array}{cccccccccccc}
   -0.0021 & -0.0019 & 4.7\times 10^{\text{-4}} & -0.0025 \\
   0.0118 & -0.0143 & 0.0118 & -1.4\times 10^{\text{-4}} \\
   0.0028 & 0.0012 & 0.0027 & -2.8\times 10^{\text{-4}} \\
   0.0029 & -0.0025 & 0.0093 & -7.5\times 10^{\text{-5}} \\
   -0.0092 & 0.0081 & -0.0066 & 0.0041 \\
   0.0081 & -0.0063 & 0.0036 & 2.1\times 10^{\text{-6}} \\
   0.0083 & -0.0022 & 0.0023 & 0.0020 \\
   -0.0178 & 0.0333 & 0.0068 & -8.3\times 10^{\text{-6}}
  \end{array}
  \right), \nonumber\\
E'^{\text{sy}} &= \left(
  \begin{array}{cccccccccccc}
   0.0031 & 0.0052 & 8.6\times 10^{\text{-4}} & 7.8\times 10^{\text{-4}} \\
   0.0047 & 0.0081 & 0.0021 & 0.0011 \\
   6.3\times 10^{\text{-4}} & 0.0036 & 0.0030 & 7.2\times 10^{\text{-4}} \\
   0.0032 & 0.0085 & 0.0065 & 3.2\times 10^{\text{-5}} \\
   0.0027 & 0.0043 & 0.0025 & 2.1\times 10^{\text{-4}} \\
   0.0015 & 0.0013 & 0.0011 & 2.7\times 10^{\text{-4}} \\
   0.0051 & 0.0045 & 0.0013 & 7.4\times 10^{\text{-4}} \\
   0.0041 & 0.0077 & 0.0055 & 6.7\times 10^{\text{-5}}
  \end{array}
  \right).
\end{align}
\end{scriptsize}

$\beta=5.40$, lattice size: $24^3\times48$
\begin{scriptsize}
\begin{align}
Z &= \left(
  \begin{array}{cccccccccccc}
   1.3340 & -0.0073 & 0.0235 & -0.0019 & -0.0065 & -0.2075 & 0.0934 & 0.0061 \\
   -0.0048 & 1.2800 & -5.5\times 10^{\text{-4}} & -1.1\times 10^{\text{-4}} & \
  -0.0925 & -0.0853 & 0.0902 & -3.2\times 10^{\text{-4}} \\
   0.0124 & 0.0037 & 1.2850 & -1.2\times 10^{\text{-4}} & -0.0060 & -0.0017 & \
  0.1101 & -4.2\times 10^{\text{-5}}  \\
   0.0059 & 0.0091 & -0.0024 & 1.3290 & 0.0014 & 0.0078 & -0.0142 & 0.1231  \\
   0.0301 & -0.0645 & 0.0426 & 8.1\times 10^{\text{-5}} & 1.3430 & -0.0877 & \
  0.0288 & 0.0013  \\
   -0.0581 & -0.0559 & 0.0530 & 2.5\times 10^{\text{-4}} & -0.0606 & 1.2850 & \
  0.0649 & -0.0015  \\
   0.0298 & 0.0263 & 0.0589 & -2.7\times 10^{\text{-4}} & -0.0027 & 0.0605 & \
  1.3380 & -3.3\times 10^{\text{-4}}  \\
   0.0047 & -4.3\times 10^{\text{-4}} & 9.7\times 10^{\text{-5}} & 0.0401 & \
  -0.0015 & 0.0043 & -0.0015 & 1.3740 
  \end{array}
  \right), \nonumber\\
E^{\text{sy}} &= \left(
  \begin{array}{cccccccccccc}
   0.0323 & 0.0107 & 0.0056 & 8.8\times 10^{\text{-4}} & 0.0103 & 0.0416 & \
  0.0375 & 9.1\times 10^{\text{-4}}  \\
   0.0120 & 0.0169 & 0.0025 & 2.7\times 10^{\text{-4}} & 0.0025 & 0.0533 & \
  0.0461 & 4.4\times 10^{\text{-4}}  \\
   0.0045 & 0.0053 & 0.0151 & 5.9\times 10^{\text{-5}} & 0.0015 & 0.0092 & \
  0.0148 & 2.0\times 10^{\text{-4}}  \\
   0.0013 & 0.0028 & 0.0016 & 0.0124 & 0.0017 & 0.0018 & 0.0030 & 0.0126  \\
   0.0145 & 0.0347 & 0.0130 & 2.7\times 10^{\text{-4}} & 0.0279 & 0.0628 & \
  0.0431 & 3.6\times 10^{\text{-4}}  \\
   0.0203 & 0.0274 & 0.0136 & 2.4\times 10^{\text{-4}} & 0.0161 & 0.0228 & \
  0.0452 & 4.2\times 10^{\text{-4}}  \\
   0.0044 & 0.0111 & 0.0092 & 3.4\times 10^{\text{-4}} & 0.0079 & 0.0133 & \
  0.0487 & 4.7\times 10^{\text{-4}}  \\
   0.0016 & 0.0013 & 0.0012 & 0.0127 & 1.7\times 10^{\text{-4}} & 0.0013 & \
  6.4\times 10^{\text{-4}} & 0.0421 
  \end{array}
  \right).
\end{align}
\begin{align}
Z' &= \left(
  \begin{array}{cccccccccccc}
   -0.0017 & -0.0021 & 5.5\times 10^{\text{-4}} & -0.0026 \\
   0.0116 & -0.0152 & 0.0114 & 1.1\times 10^{\text{-4}} \\
   0.0025 & -4.2\times 10^{\text{-5}} & 0.0034 & -9.0\times 10^{\text{-6}} \\
   0.0028 & -1.1\times 10^{\text{-4}} & 0.0068 & -7.7\times 10^{\text{-5}} \\
   -0.0089 & 0.0085 & -0.0066 & 0.0040 \\
   0.0074 & -0.0057 & 0.0037 & -1.0\times 10^{\text{-4}} \\
   0.0083 & -0.0028 & 0.0025 & 0.0021 \\
   -0.0158 & 0.0321 & 0.0059 & -1.3\times 10^{\text{-5}}
  \end{array}
  \right), \nonumber\\
E'^{\text{sy}} &= \left(
  \begin{array}{cccccccccccc}
   0.0018 & 0.0019 & 4.8\times 10^{\text{-4}} & 6.3\times 10^{\text{-4}} \\
   0.0028 & 0.0060 & 0.0011 & 8.7\times 10^{\text{-4}} \\
   7.4\times 10^{\text{-4}} & 0.0035 & 0.0027 & 7.2\times 10^{\text{-4}} \\
   0.0016 & 0.0090 & 0.0065 & 2.6\times 10^{\text{-5}} \\
   0.0014 & 0.0029 & 0.0016 & 1.4\times 10^{\text{-4}} \\
   0.0020 & 0.0016 & 9.0\times 10^{\text{-4}} & 3.7\times 10^{\text{-4}} \\
   0.0031 & 0.0032 & 0.0012 & 4.9\times 10^{\text{-4}} \\
   0.0043 & 0.0043 & 0.0036 & 1.8\times 10^{\text{-5}}
  \end{array}
  \right).
\end{align}
\end{scriptsize}

\end{document}